\documentclass[prd,aps,showpacs,preprintnumbers,amsmath,amssymb,onecolumn,superscriptaddress,prl,floatfix,nofootinbib]{revtex4-2}

\usepackage{amsfonts}
\usepackage{amsmath}
\usepackage{amssymb}
\usepackage{graphicx} 
\usepackage{dcolumn}  
\usepackage{bm,bbm}   
\usepackage{tabularx}
\usepackage{epstopdf}
\usepackage{xkcdcolors}
\usepackage{xcolor}

\definecolor{china}{rgb}{0.0, 0.42, 0.24}
\definecolor{west}{rgb}{0.19, 0.55, 0.91}
\definecolor{japan}{rgb}{0.0, 0.87, 0.87}
\definecolor{asean}{rgb}{1.0, 0.03, 0.0}
\definecolor{russia}{rgb}{0.97, 0.5, 0.75}
\definecolor{india}{rgb}{1.0, 0.49, 0.0}
\definecolor{jv}{rgb}{0.57, 0.64, 0.69}
\definecolor{southam}{rgb}{0.58, 0.34, 0.92}
\definecolor{notinproj}{rgb}{1.0, 0.75, 0.0}
\definecolor{africa}{HTML}{964b00}
\definecolor{middleeast}{HTML}{00008b}

\definecolor{gp}{rgb}{0.58, 0.34, 0.92}
\definecolor{el}{rgb}{0.19, 0.55, 0.91}
\definecolor{pw}{rgb}{1.0, 0.03, 0.0}
\definecolor{cb}{rgb}{0.2, 0.8, 0.2}
\definecolor{ie}{rgb}{1.0, 0.49, 0.0}

\usepackage{mathrsfs}
\usepackage{subcaption}
\usepackage{mathtools}
\usepackage{float}
\usepackage[linesnumbered,lined,boxed,commentsnumbered,ruled,vlined]{algorithm2e}
\usepackage{algpseudocode}
\usepackage{verbatim}
\usepackage{comment}
\usepackage{ragged2e}
\usepackage{physics}
\usepackage{hyperref} 
\usepackage{url}
\usepackage{cleveref}
\usepackage{tikz}

\makeatletter
\gdef\@ptsize{1} 
\makeatother
\usepackage{setspace}
\doublespace

\DeclareCaptionJustification{justified}{\justifying}
\hypersetup{colorlinks=true, citecolor=orange, urlcolor=blue, linkcolor=magenta}

\begin{document}

\title{Pattern-detection in the global automotive industry:\\a manufacturer-supplier-product network analysis}

\author{Massimiliano Fessina}
\email{massimiliano.fessina@imtlucca.it}
\affiliation{IMT School for Advanced Studies, Piazza San Francesco 19, 55100 Lucca (Italy)}
\author{Andrea Zaccaria}
\affiliation{Institute for Complex Systems (ISC), National Research Council, Piazzale Aldo Moro 2, 00185 Rome (Italy)}
\affiliation{Enrico Fermi Research Center (CREF), Via Panisperna 89A, 00184 Rome (Italy)}
\author{Giulio Cimini}
\affiliation{Physics Department and INFN, University of Rome Tor Vergata, Via della Ricerca Scientifica 1, 00133 Rome (Italy)}
\affiliation{Enrico Fermi Research Center (CREF), Via Panisperna 89A, 00184 Rome (Italy)}
\author{Tiziano Squartini}
\affiliation{IMT School for Advanced Studies, Piazza San Francesco 19, 55100 Lucca (Italy)}

\date{\today}
 
\begin{abstract}
Production networks arise from customer-supplier relationships between firms. These systems have gained increasing attention as a consequence of the frequent supply chain disruptions caused by the natural and man-made disasters occurred during the last years (e.g. the Covid-19 pandemic and the Russia-Ukraine war). Recent, empirical evidence has shown that production networks are shaped by `functional' structures reflecting the complementarity of firms, i.e. their tendency to compete. However, data constraints force the few, available studies to consider only country-specific production networks. In order to fully capture the cross-country structure of modern supply chains, here we focus on the global, automotive industry as depicted by the `MarkLines Automotive' dataset. After representing it as a network of manufacturers, suppliers and products, we look for the statistical significance of the aforementioned, `functional' structures. Our exercise reveals the presence of several pairs of manufacturers sharing a significantly large number of suppliers, a result confirming that any two car companies are seldom engaged in a buyer-supplier relationship: rather, they compete although being connected to many, common neighbors. Interestingly, `generalist' suppliers serving many manufacturers co-exist with `specialist' suppliers serving few manufacturers. Additionally, we unveil the presence of patterns with a clearly spatial signature, with manufacturers clustering around groups of geographically close suppliers: for instance, Chinese firms constitute a disconnected community, likely an effect of the protectionist policies promoted by the Chinese government. We also show the tendency of suppliers to organize their production by targeting specific car systems, i.e. combinations of technological devices designed for specific tasks. Besides shedding light on the self-organizing principles shaping production networks, our findings open up the possibility of designing realistic generative models of supply chains, to be used for testing the resilience of the global economy.
\end{abstract}

\maketitle

\section{Introduction}

The growth of network science over the last twenty years has impacted several disciplines, by establishing new, empirical facts about the structural properties of those systems as well as novel methodologies for their analysis. Prominent examples are represented by economic and financial networks, such as the international trade~\cite{serrano2003topology,garlaschelli2005structure,saracco2015randomizing}, product exports by countries~\cite{Hidalgo:2009aa,Tacchella:2018aa,Pugliese:2019aa,balland2022new}, transaction networks~\cite{lin2020lightning,mattsson2023circulation} and, after the 2008 crisis, interbank networks~\cite{boss2004network,battiston2012debtrank,squartini2013early,cimini2015systemic,bargigli2015multiplex}. A class of systems that has recently gained attention is that of production networks emerging as (manufacturing) firms become dependent on other (supplier) firms for their own production. As a consequence of globalization~\cite{Chase-Dunn:2000aa} and of a constant strive towards efficiency~\cite{Craighead:2007aa}, production networks have become increasingly interdependent - a feature lying at the basis of the business interruptions that have occurred in consequence of the recent, natural and man-made disasters (e.g. the Covid-19 pandemic and the Russia-Ukraine war)~\cite{Inoue:2019aa,guan2020global,Aldrighetti:2021aa,Carvalho:2021aa,Ivanov:2021aa,chowdhury2021covid}.

The propagation of economic shocks has been traditionally studied using \emph{industry-level} input-output tables~\cite{Miller:2009aa}, although the importance of \emph{firm-level} supply data to provide an unbiased assessment of the consequences of individual failures~\cite{Morimoto:1970aa,diem2023estimating} has been increasingly recognized~\cite{Bak:1993aa,Gabaix:2011aa,Acemoglu:2012aa}. Network theory is, thus, becoming an increasingly popular tool to analyze production systems~\cite{Choi:2001aa,Surana:2005aa,Wycisk:2008aa,Kim:2015aa,Brintrup:2017aa,Perera:2017aa,Brintrup:2018aa,Barrot:2016aa,Demirel:2019aa,Demir:2022aa,diem2022quantifying,fujiwara2016debtrank}, analogously to what has happened within the realm of financial systems in the aftermath of the 2008 crisis~\cite{bardoscia2021physics}.

Firm-level datasets, however, are notoriously difficult to acquire because of both technical and privacy issues~\cite{bacilieri2022firm}. While early works relied on aggregated flows of goods between countries~\cite{Lee:2011aa,Mizuno:2016aa,Gephart:2016aa,Klimek:2019aa,Starnini:2019aa}, more detailed datasets have recently appeared: still, the number of firm-level datasets, with large-scale coverage, that are built from financial reports is limited; besides, they only cover the larger companies listed on the US stock exchanges and their main customers. This is the case of Factset~\cite{Konig:2022aa,bacilieri2022firm}, Compustat~\cite{Atalay:2011aa,Cohen:2008aa,Barrot:2016aa} and Capital IQ~\cite{Chakraborty:2020aa,chakraborty2021inequality} data. Other, global datasets that have been analyzed in the Operations Research and Supply Chain Management literature cover specific, industrial sectors~\cite{Wiedmer:2021aa}: the most popular, and complete, one concerns the automotive sector, obtained from a private industry database (the `MarkLines Automotive Information Platform') compiled through surveys sent to automotive supplier firms~\cite{Brintrup:2015aa,brintrup2018predicting}. At the level of individual countries, instead, production networks can be constructed either from value-added tax (VAT) data (which exists for Belgium~\cite{dyne2015belgian}, Ecuador~\cite{Mungo:2023aa}, Hungary~\cite{diem2022quantifying} and Spain~\cite{peydro2020production}) or from payment data mediated by central, or major, banks (which exists for Brazil~\cite{silva2020brasil}, Japan~\cite{tamura2012estimation,Fujiwara2021regional} and The Netherlands~\cite{ialongo2022reconstructing}). National data, often characterized by a reporting threshold, typically provide a very good internal coverage.

The Japanese, inter-firm dataset provided by Tokyo Shoko Research (TSR) was the first, national, firm-level network to become available and, as such, has been extensively studied during the last decade~\cite{saito2007larger,ohnishi2009hubs,ohnishi2010network,fujiwara2010large}. It provides credit information of about one million firms, as well as the identity of the most relevant suppliers and customers for the business activity of each. Empirical analyses revealed that both firm-specific, structural quantities (e.g. the number of connections) and purely financial indicators (e.g. the total amount of sales, the total number of employees) are power-law distributed, while a firm degree grows along with its total sales in a non-linear fashion \cite{saito2007larger}. Regarding the topology of the network, it was shown to be disassortative by degree (as suppliers with few customers are preferentially connected with large firms) and characterized by a well-defined community structure, with clusters being shaped by geographical proximity and industrial sector similarity~\cite{fujiwara2010large}. Notably, these clusters are characterized by bipartite structures with large companies not being directly linked but sharing many first-tier suppliers. This result was confirmed by a comprehensive study of triadic motifs~\cite{ohnishi2010network}, i.e. all, possible interaction patterns involving three nodes~\cite{alonuri2007}. When compared with a benchmark constraining the degree of each firm, the TSR data features an under-representation of closed triangles and an over-representation of V-motifs~\cite{saracco2015randomizing}, i.e. the patterns shaped by pairs of nodes (firms) that are not directly connected but share a common neighbor (be it a supplier or a customer).

The abundance of V-motifs in production networks can be interpreted as a sign that their self-organization is driven by \emph{complementarity} (`similar nodes tend to share the same partners') rather than \emph{homophily} (`similar nodes tend to be connected')~\cite{mattsson2021functional}. From this perspective, production networks are more similar to protein-interaction networks than to social networks: in the first case, two proteins with similar binding sites do not interact directly but can be both connected with others having complementary binding properties~\cite{Kovacs:2019aa}; in social networks, on the contrary, acquaintances, people with similar interests, etc. are likely to be connected (according to the motto `the friend of my friend is my friend'~\cite{Liben-Nowell:2007aa}). Additionally, the over-representation of V-motifs is compatible with the over-representation of the square patterns known as X-motifs~\cite{saracco2015randomizing}, given by two nodes (e.g. firms) that are not directly connected but share pairs of common neighbors (be they suppliers or customers). The findings of~\cite{fujiwara2010large,ohnishi2010network}, thus, corroborate the picture according to which firms with similar outputs are often engaged in a competition, rather than in a buyer-supplier relationship - although they can have many suppliers and customers in common~\cite{brintrup2018predicting}.

The aforementioned, over-representation of square motifs, however, has been explicitly shown only for the company-level, production network provided by CBS - Statistics Netherlands~\cite{CBS2019,mattsson2021functional}: no results are available for global datasets, despite the well-documented cross-country structure of production networks~\cite{antras2022global}. The aim of this paper is bridging the gap, by carrying out a pattern-detection analysis on production networks at the global scale, employing the `MarkLines Automotive' dataset. In particular, after representing the system in terms of a manufacturer-supplier-product network, we carry out a preliminary description of its structural features. Then, we test their statistical relevance by adopting a validation framework based on the maximum-entropy principle. Specifically, we apply the method proposed in~\cite{saracco2017inferring}, which takes as input the bipartite, manufacturer-supplier representation of the `MarkLines Automotive' dataset and produces monopartite projections whose connections are induced by the similarity of the corresponding nodes - in our case, the number of V-motifs, i.e. the common neighbors in the original, bipartite network. In statistical terms, this amounts to unveiling the presence of significant patterns as a consequence of the rejection of the null hypothesis that the node degrees are the only explanatory variables of the network~\cite{saracco2015randomizing,Gualdi:2016aa,saracco2017inferring,vallarano2021fast}; filtering out the statistical noise allows us to discard the spurious patterns (e.g. the motifs emerging as a simple consequence of the node degrees~\cite{saracco2015randomizing}) while retaining those emerging as a genuine consequence of the interactions between nodes. Remarkably, the different kinds of structures we detect seem to clearly reflect national development strategies.

\begin{figure*}[t!]
\centering 
\includegraphics[width=\textwidth]{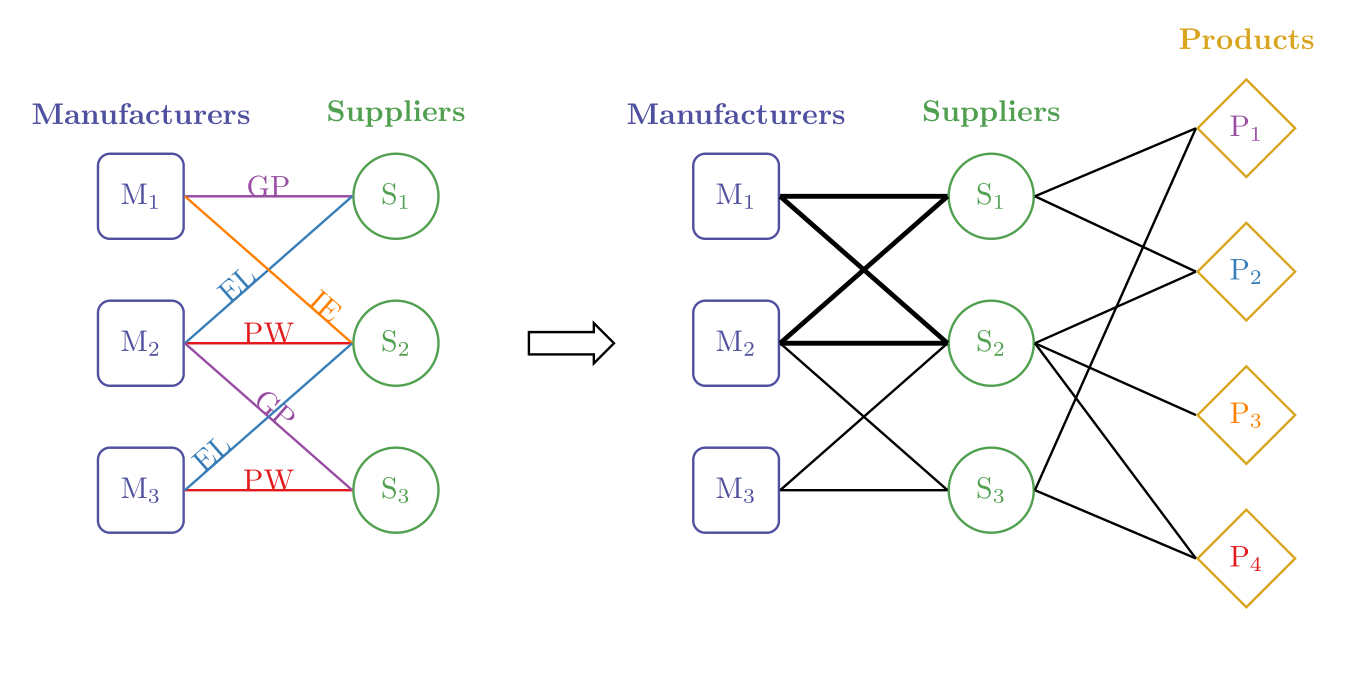}
\caption{The `MarkLines Automotive' dataset, represented either as a bipartite multiplex (left) or as a tripartite network (right). In the multiplex representation, there are five layers, corresponding to the product categories of the data: a link between a manufacturer and a supplier corresponds to the exchange of a specific product, belonging to the corresponding category and colored accordingly. In the tripartite representation, a link between a manufacturer and a supplier is present as long as they are connected by, at least, one link across all layers; each supplier, instead, is connected to all the products it sells. Bold lines illustrate the motifs proxying competition between manufacturers and that we statistically validate.}
\label{fig1}
\end{figure*}

\section{Methods}

\subsection{Network representation of the `MarkLines Automotive' dataset}

The `MarkLines Automotive' dataset (\url{https://www.marklines.com}) identifies the `manufacturers' with the companies making cars (e.g. \emph{Honda}) and the `suppliers' with the companies selling them products (e.g. \emph{Hitachi}). After a  thorough data-cleaning and harmonization procedure (see Appendix A), we consider $M=301$ manufacturers, $S=5.725$ suppliers, and $P=291$ products, while the overall number of buyer-supplier relationships amounts to $26.535$ (see Table~\ref{tab1}). As there are few, `internal' links between manufacturers (a car maker can happen to act as a supplier of another car maker) and no links between suppliers, the dataset admits a natural, bipartite representation once the `internal' links are discarded.

\begin{table}[b!]
\caption{\label{tab1}Basic statistics of the `MarkLines Automotive' dataset for each category of products: Chassis/Body (CB), Electrical (EL), Powertrain (PW), Interior/Exterior (IE) and General parts (GP). The percentages of `non-bipartite' links refer to the connections between manufacturers for each product category.}
\begin{ruledtabular}
\begin{tabular}{lllllll}
                 & \textbf{CB} & \textbf{EL} & \textbf{PW} & \textbf{IE} & \textbf{GP} & \textbf{Full}\\
\hline
\# manufacturers & 246  & 215  & 249   & 196  & 108  & 301\\
\# suppliers     & 1.864 & 1.203 & 2.659  & 1.776 & 479  & 5.725\\
\# links         & 7.937 & 5.390 & 10.978 & 6.502 & 1.687 & 26.535\\
\hline
\% `non-bipartite' links  & 0.4 & 0.94 & 2.6 & 0.83 & 0 & 1.3\\
\end{tabular}
\end{ruledtabular}
\end{table}

According to the (technological) taxonomy provided by the platform, products are classified into five categories: Chassis/Body (CB), Electrical (EL), Powertrain (PW), Interior/Exterior (IE) and General parts (GP). Information about each category of products, indexed by $\gamma=1\dots5$, can be arranged into a biadjacency matrix $\mathbf{B}^\gamma$ whose dimensions read $M^\gamma\times S^\gamma$, where $M^\gamma$ is the number of manufacturers and $S^\gamma$ is the number of suppliers in category $\gamma$: naturally, $b_{ms}^\gamma=1$ if supplier $s$ provides manufacturer $m$ with at least one product belonging to category $\gamma$ and $b_{ms}^\gamma=0$ otherwise. Overall, then, the `MarkLines Automotive' dataset can be represented as a bipartite multiplex, each layer corresponding to a category of technological products (see also the left panel of Figure~\ref{fig1}). 

Alternatively, the dataset can be represented as an $M\times S\times P$ tripartite network, i.e. a `combination' of two bipartite networks sharing the set of suppliers (see also the right panel of Figure \ref{fig1}): the generic element of the `left' $M\times S$ biadjacency matrix $\mathbf{L}$ reads $l_{ms}=1$ if manufacturer $m$ buys from supplier $s$, otherwise $l_{ms}=0$; analogously, the generic element of the `right' $S\times P$ biadjacency matrix $\mathbf{R}$ reads $r_{sp}=1$ if supplier $s$ sells product $p$ and $r_{sp}=0$ otherwise.
Although we could also link each manufacturer to the set of products purchased by its suppliers, the evidence that all of them are car producers, hence needing the same basket of products, let us opt for discarding this third set of connections\footnote{Intuition would suggest each manufacturer to need a minimum amount of products to make a car. This, however, does not seem to be the case as manufacturers exist that are linked to just one supplier but no supplier exists that provides all products. This, in turn, suggests that the `MarkLines Automotive' dataset is (at least, partially) incomplete.}. Hereby, we will adopt the tripartite representation.

\subsection{Structural properties}

\paragraph*{Local connectivity.} The most important network quantity at the local level is the degree, defined as the number of connections of a node. In order to properly describe our data, we need the following definitions:

\begin{itemize}
\item the number of suppliers, or product providers, of manufacturer $m$: $k_{m\rightarrow\mathcal{S}}=\sum_{s=1}^Sl_{ms}$;
\item the number of customers, or client manufacturers, of supplier $s$: $k_{s\rightarrow\mathcal{M}}=\sum_{m=1}^Ml_{ms}$;
\item the `diversification' of supplier $s$, i.e. the number of products it sells to its client manufacturers: $k_{s\rightarrow\mathcal{P}}=\sum_{p=1}^P r_{sp}$;
\item the `ubiquity' of product $p$, i.e. the number of its vendor suppliers: $k_{p\rightarrow\mathcal{S}}=\sum_{s=1}^S r_{sp}$.
\end{itemize}

\paragraph*{Assortativity.} The presence of degree correlations is captured by the average nearest neighbors degree. In order to properly describe our data, we need the following definitions:

\begin{itemize}
\item the average number of customers of a manufacturer neighbors: $\text{ANND}_{m\rightarrow\mathcal{S}}=\sum_{s=1}^Sl_{ms}k_{s\rightarrow\mathcal{M}}/k_{m\rightarrow\mathcal{S}}$;
\item the average number of providers of a supplier neighbors: $\text{ANND}_{s\rightarrow\mathcal{M}}=\sum_{m=1}^Ml_{ms}k_{m\rightarrow\mathcal{S}}/k_{s\rightarrow\mathcal{M}}$;
\item the average `ubiquity' of a supplier products: $\text{ANND}_{s\rightarrow\mathcal{P}}=\sum_{p=1}^Pr_{sp}k_{p\rightarrow\mathcal{S}}/k_{s\rightarrow\mathcal{P}}$;
\item the average `diversification' of a product suppliers: $\text{ANND}_{p\rightarrow\mathcal{S}}=\sum_{s=1}^Sr_{sp}k_{s\rightarrow\mathcal{P}}/k_{p\rightarrow\mathcal{S}}$.
\end{itemize}

\paragraph*{Motifs.} In order to capture the concept of a node `highly connected neighborhood', we need to consider the bipartite analogue of the monopartite, square clustering coefficient. To this aim, several definitions have been proposed so far (see also Appendix B): here, we adopt the one provided in~\cite{zhang2008clustering} and reading\footnote{In order to keep the discussion as simple as possible, we provide the definition of the bipartite clustering coefficient only for manufacturers. Whereas needed, it will be extended to suppliers and products.}

\begin{equation}
\text{BCC}_m=\frac{\sum_{s<s'}q_m(s,s')}{\sum_{s<s'}[(k_{s\rightarrow\mathcal{M}}-1)+(k_{s'\rightarrow\mathcal{M}}-1)-q_{m}(s,s')]}
\end{equation}
where $k_{s\rightarrow\mathcal{M}}$ is the number of customers, or client manufacturers, of supplier $s$ and

\begin{equation}
q_m(s,s')=\sum_{\substack{m'=1\\(m'\neq m)}}^Ml_{ms}l_{ms'}l_{m's}l_{m's'}=l_{ms}l_{ms'}\sum_{\substack{m'=1\\(m'\neq m)}}^Ml_{m's}l_{m's'}
\end{equation}
counts the number of cycles, composed by four links, involving the common neighbors of $s$ and $s'$, other than $m$. According to $\text{BCC}_m$, the total number of closed squares involving manufacturer $m$ is given by the sum of the degrees of all pairs of its neighbors minus $q_{m}(s,s')$, i.e. the number of squares that are already closed. In other words, the total number of closed squares coincides with the number of squares that could become closed upon connecting, by adding \emph{new} links, the neighbors of $s$ with $s'$ and the neighbors of $s'$ with $s$, excluding $m$ from both sets.

Upon considering that the generic addendum $l_{ms}l_{ms'}l_{m's}l_{m's'}$ equals 1 if a closed square involving nodes $m$, $m'$, $s$ and $s'$ exists and 0 otherwise, we can compute the number of squares involving manufacturers $m$ and $m'$ by summing over $s$ and $s'$, i.e. as

\begin{equation}\label{xmot}
X_{mm'}=\sum_{s<s'}l_{ms}l_{ms'}l_{m's}l_{m's'}=\binom{V_{mm'}}{2}
\end{equation}
where $V_{mm'}\equiv\sum_{s=1}^Sl_{ms}l_{m's}$ is the number of common suppliers (V-motifs) to $m$ and $m'$~\cite{saracco2015randomizing}. Hence, the number of cycles, composed by four links, involving node $m$ equals the number of X-motifs involving it, i.e. $\sum_{s<s'}q_m(s,s')=\sum_{\substack{m'=1\\(m'\neq m)}}^MX_{mm'}$.\\

\paragraph*{Subgraph centrality.} In order to detect the presence of closed paths of higher order, we have considered the bipartite analogue of the monopartite subgraph centrality~\cite{estrada2005subgraph}, whose definition reads

\begin{equation}
\text{BSC}_m=\frac{\sum_{k=0}^{\infty}[\mathbf{L}^{2k}]_{mm}}{(2k)!}=[\text{cosh}(\mathbf{L})]_{mm}
\end{equation}
where $\mathbf{L}^0\equiv\mathbf{I}$, i.e. the zero-th power of $\mathbf{L}$ coincides with the $M\times M$ identity matrix, and only closed walks of even length are accounted for. Still, carrying out a meaningful comparison of the centrality of nodes across different configurations requires it to be properly normalized; hereby, we adopt the following definition

\begin{equation}
\overline{\text{BSC}}_m=\frac{\text{BSC}_m}{\sum_{m=1}^M\text{BSC}_m}=\frac{[\text{cosh}(\mathbf{L})]_{mm}}{\text{Tr}[\text{cosh}(\mathbf{L})]}.
\end{equation}

\paragraph*{Nestedness.} Finally, we inspect the mesoscale organization of our bipartite networks by measuring their (degree of) nestedness~\cite{MARIANI20191}, i.e. the tendency of low-degree nodes to have neighbors that are a subset of the neighbors of high-degree nodes. This pattern leads to a `triangular' structure that has been observed in ecological systems~\cite{bascompte2003}, economic systems~\cite{tacchella2012new}, etc. To quantify (the degree of) nestedness, we employ the NODF~\cite{almeida2008consistent}, reading

\begin{equation}
\text{NODF(L)}=\frac{1}{K}\left[\sum_{m,m'=1}^{M}\left(\theta(k_{m\rightarrow \mathcal{S}}-k_{m'\rightarrow \mathcal{S}})\frac{\sum_{s=1}^{S}l_{ms}l_{m's}}{k_{m'\rightarrow \mathcal{S}}}\right)+\sum_{s,s'=1}^{S}\left(\theta(k_{s\rightarrow \mathcal{M}}-k_{s'\rightarrow \mathcal{M}})\frac{\sum_{m=1}^{M}l_{ms}l_{ms'}}{k_{s'\rightarrow \mathcal{M}}}\right)\right].
\end{equation}

\subsection{Statistical benchmarks}

Answering the question of whether an empirical property represents a non-trivial signature of a network requires comparing it with a properly defined benchmark (or \emph{null model}, as its role is analogous to the one played by a null hypothesis in traditional statistics).

Following the approach introduced in~\cite{park2004statistical} and developed in~\cite{squartini2011analytical}, here we adopt the Exponential Random Graphs (ERG) formalism, characterizing maximum-entropy probability distributions that preserve a desired set of constraints on average while keeping everything else as random as possible~\cite{cimini2019statistical}. Among the models that can be defined within this framework, we consider the Bipartite Configuration Model (BiCM), introduced in~\cite{saracco2015randomizing} and defined by constraining the degrees of the nodes belonging to both layers of a bipartite network. In other words, this model embodies the null hypothesis that (the numerical values of) empirical network patterns are induced by (the numerical values of) the degrees of the nodes. 

Let us, now, briefly illustrate the basic results for our manufacturer-supplier network, redirecting the interested reader to~\cite{saracco2015randomizing} for the explicit derivation of the BiCM. In the case of the BiCM, constrained entropy-maximization leads to a factorized probability reading $P(\mathbf{L})=\prod_{m,s}p_{ms}^{l_{ms}}(1-p_{ms})^{1-l_{ms}}$ where

\begin{equation}
p_{ms}=\frac{x_my_s}{1+x_my_s}
\end{equation}
is the probability that manufacturer $m$ and supplier $s$ are connected (i.e. that $l_{ms}=1$) and $x_m$ and $y_s$ are functions of the Lagrange multipliers, respectively controlling for the degrees $k_{m\rightarrow S}$ and $k_{s\rightarrow M}$. These parameters, sometimes called `fitnesses', are usually related to the (economic, financial, etc.) importance of a node, as degrees are typically proportional to fluxes (export/import, money lent/money borrowed, etc.)~\cite{cimini2015systemic}.

The vectors of parameters $\boldsymbol{x}$ and $\boldsymbol{y}$ must be numerically determined. Here, we employ the maximum-likelihood principle, prescribing to maximize the expression $\mathscr{L}=\ln P(\mathbf{L}|\boldsymbol{x},\boldsymbol{y})$ with respect to $x_m$, $\forall\:m$ and $y_s$, $\forall\:s$. Such a recipe leads us to find the system of equations

\begin{align}
k_{m\rightarrow S}^*&=\sum_{s=1}^S\frac{x_my_s}{1+x_my_s},\:\forall\:m\\
k_{s\rightarrow M}^*&=\sum_{m=1}^M\frac{x_my_s}{1+x_my_s},\:\forall\:s
\end{align}
ensuring that the empirical value of each constraint matches its expectation. The system above has been solved by running the NEMTROPY package~\cite{vallarano2021fast} available at \url{https://github.com/nicoloval/NEMtropy}.

\subsection{Projection of the `MarkLines Automotive' dataset}

The projection of a bipartite network onto one of its layers yields a monopartite graph, where any two nodes are connected as the number of common neighbors, proxying their similarity~\cite{Zhou:2007aa,Cimini:2022aa}, is significantly large. Here, we follow the approach proposed in~\cite{Gualdi:2016aa,saracco2017inferring} to validate the similarity of any two nodes with respect to the BiCM. Schematically, this method works by A) focusing on a specific pair of nodes belonging to the layer of interest and counting the number of common neighbors; B) quantifying its statistical significance with respect to the BiCM; C) linking the two nodes if, and only if, the corresponding p-value is sufficiently low. Let us, now, describe these steps in detail.\\

\paragraph*{Quantifying nodes similarity.} The simplest indicator of the similarity of two nodes belonging to the same layer of a bipartite network is provided by the number of their common neighbors. For the couple of manufacturers $m$ and $m'$ this number is given by

\begin{equation}
V_{mm'}^*=\sum_{s=1}^Sl_{ms}l_{m's},
\end{equation}
i.e. the number of V-motifs the two nodes originate (as $m$ and $m'$ cannot be directly connected, the presence of a common supplier, belonging to the opposite layer, draws a V-like shape~\cite{saracco2017inferring}).\\

\paragraph*{Statistical significance of nodes similarity.} The BiCM, as any ERG model induced by linear constraints, treats links as independent random variables. Therefore, the presence of a common supplier for any two manufacturers $m$ and $m'$, i.e. $l_{ms}l_{m's}=1$, can be described as the outcome of a Bernoulli trial whose probability coefficients read

\begin{align}
&f_\text{Ber}(l_{ms}l_{m's})=p_{ms}p_{m's},\\
&f_\text{Ber}(l_{ms}l_{m's}=0)=1-p_{ms}p_{m's}.
\end{align}

As $V_{mm'}$ is a sum of independent Bernoulli trials, each characterized by a different probability, the behavior of such a random variable is described by the Poisson-Binomial (PB) distribution~\cite{saracco2017inferring}. Evaluating the statistical significance of the similarity of nodes $m$ and $m'$, thus, amounts to computing the p-value

\begin{equation}
\text{p-value}(V_{mm'})=\sum_{x\geq V_{mm'}^{*}} f_\text{PB}(x).
\end{equation}

\begin{figure*}[t!]
\centering
\textbf{A}\includegraphics[width=0.48\textwidth]{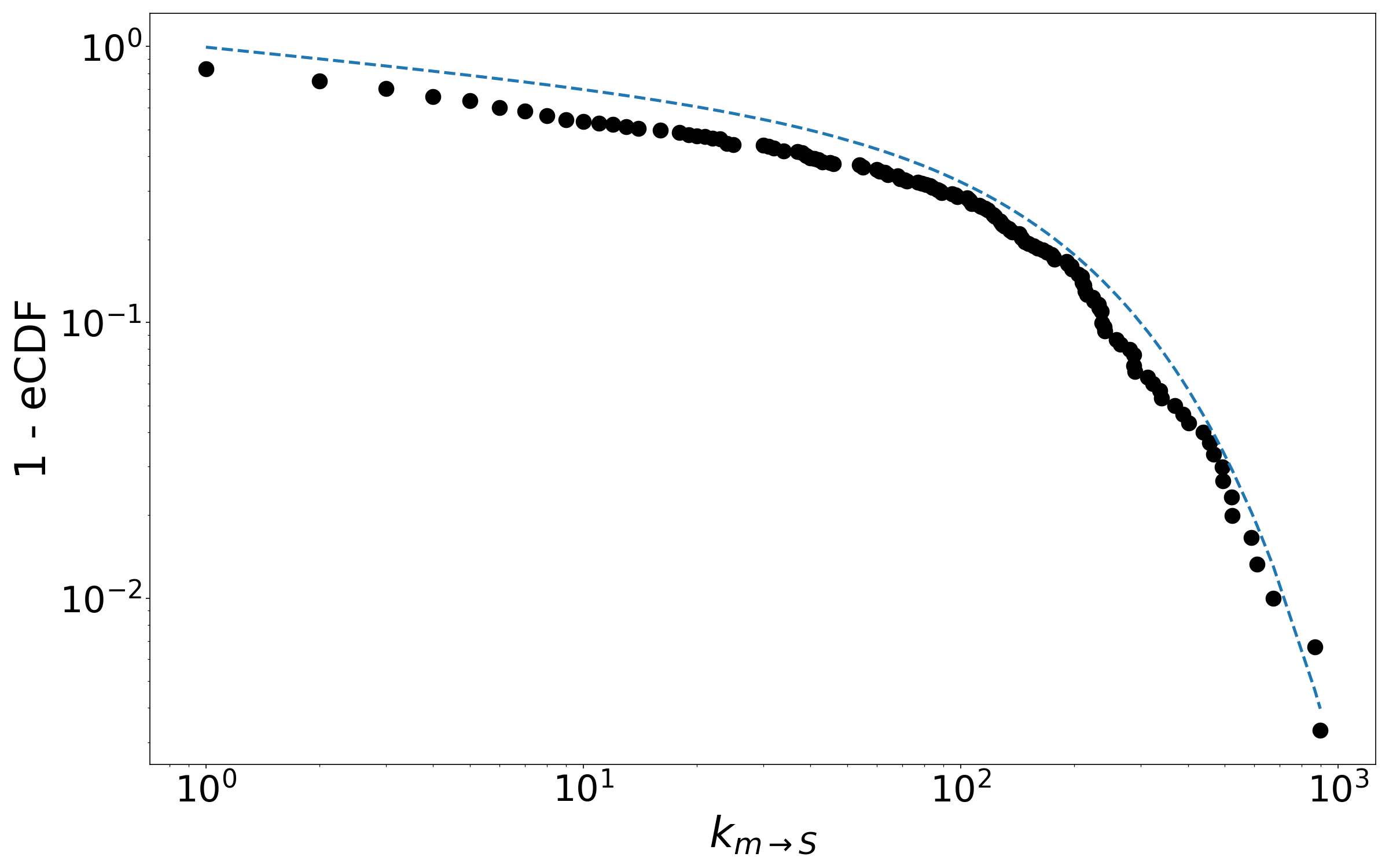}
\textbf{B}\includegraphics[width=0.48\textwidth]{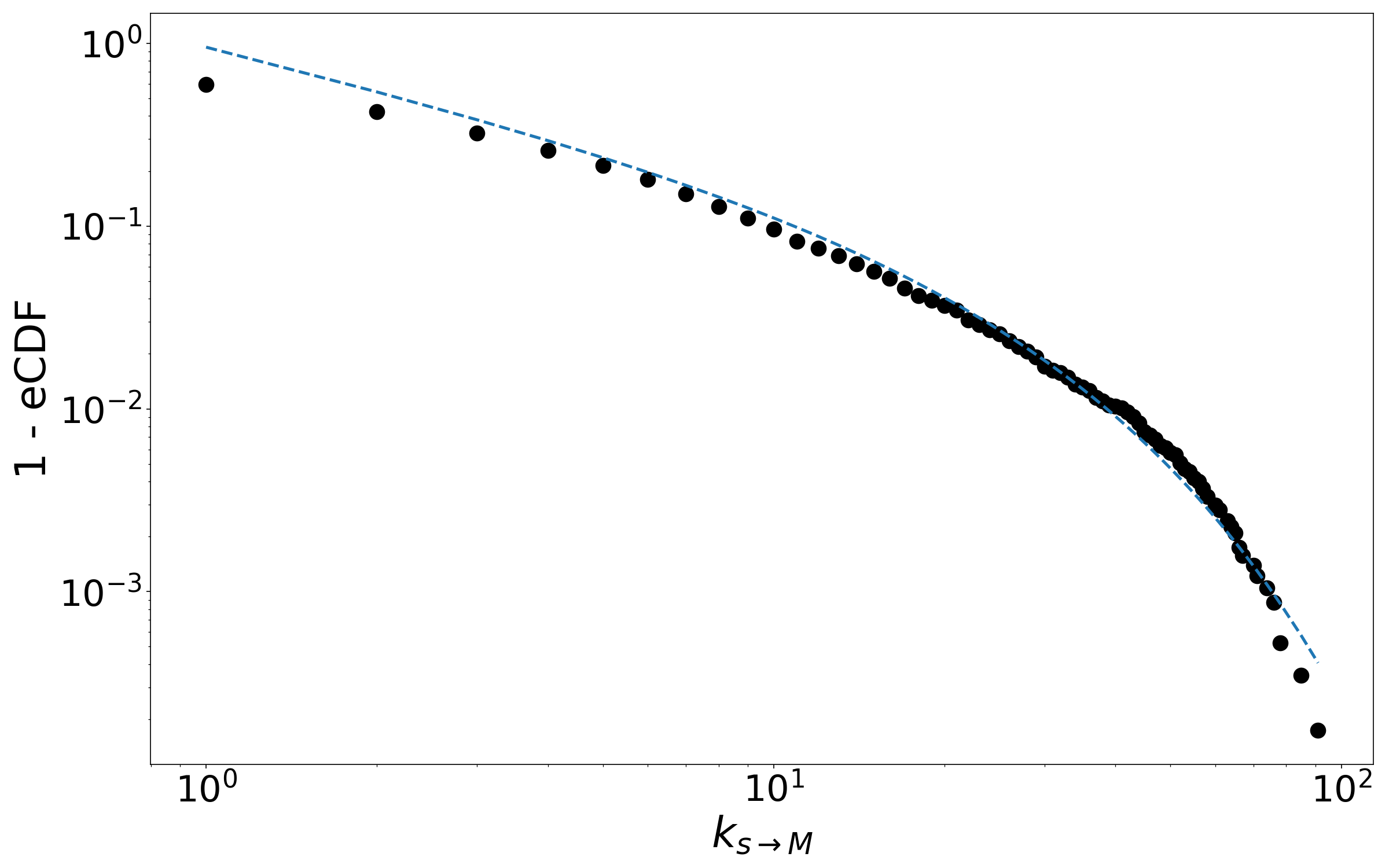}\\
\textbf{C}\includegraphics[width=0.48\textwidth]{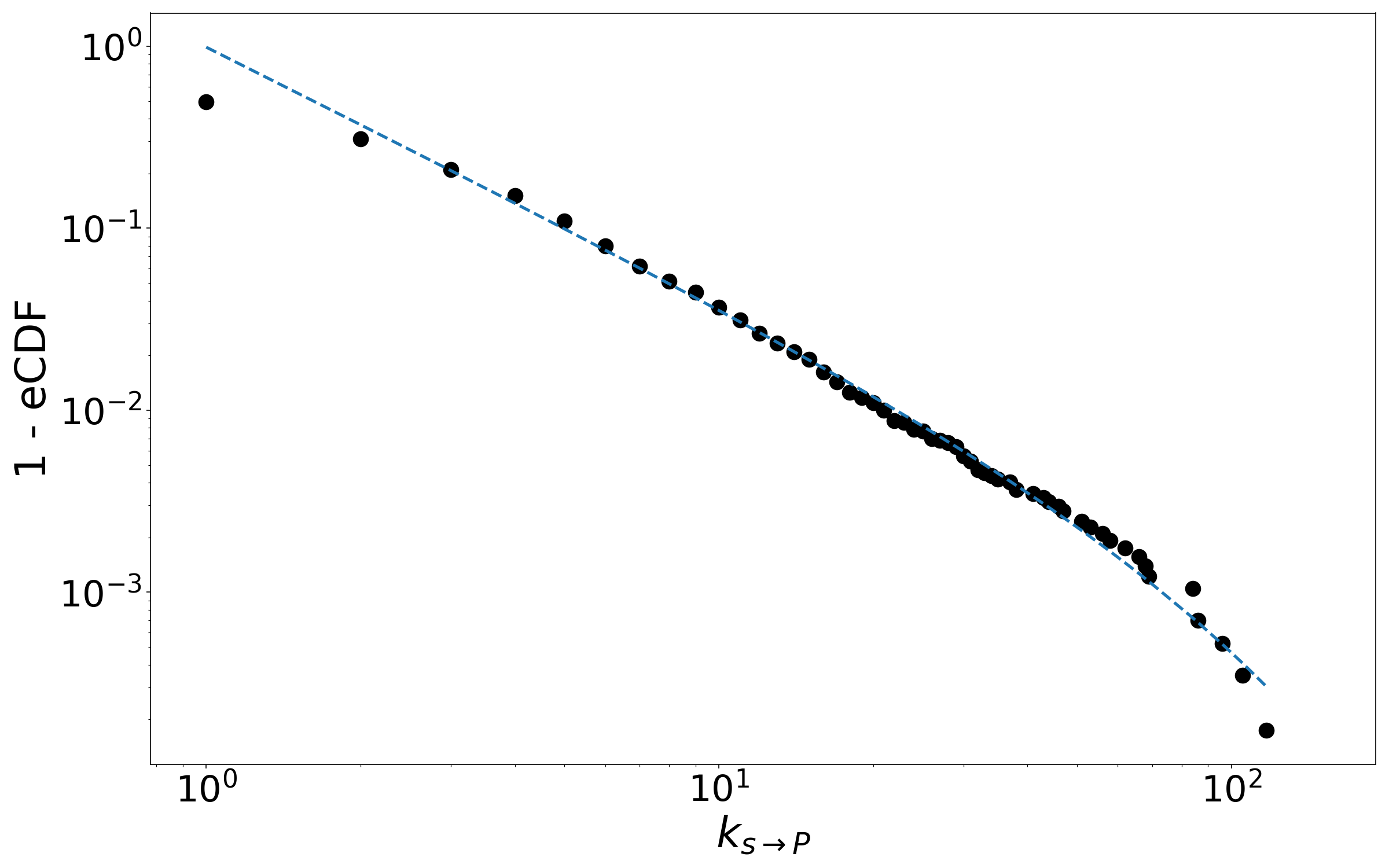}
\textbf{C}\includegraphics[width=0.48\textwidth]{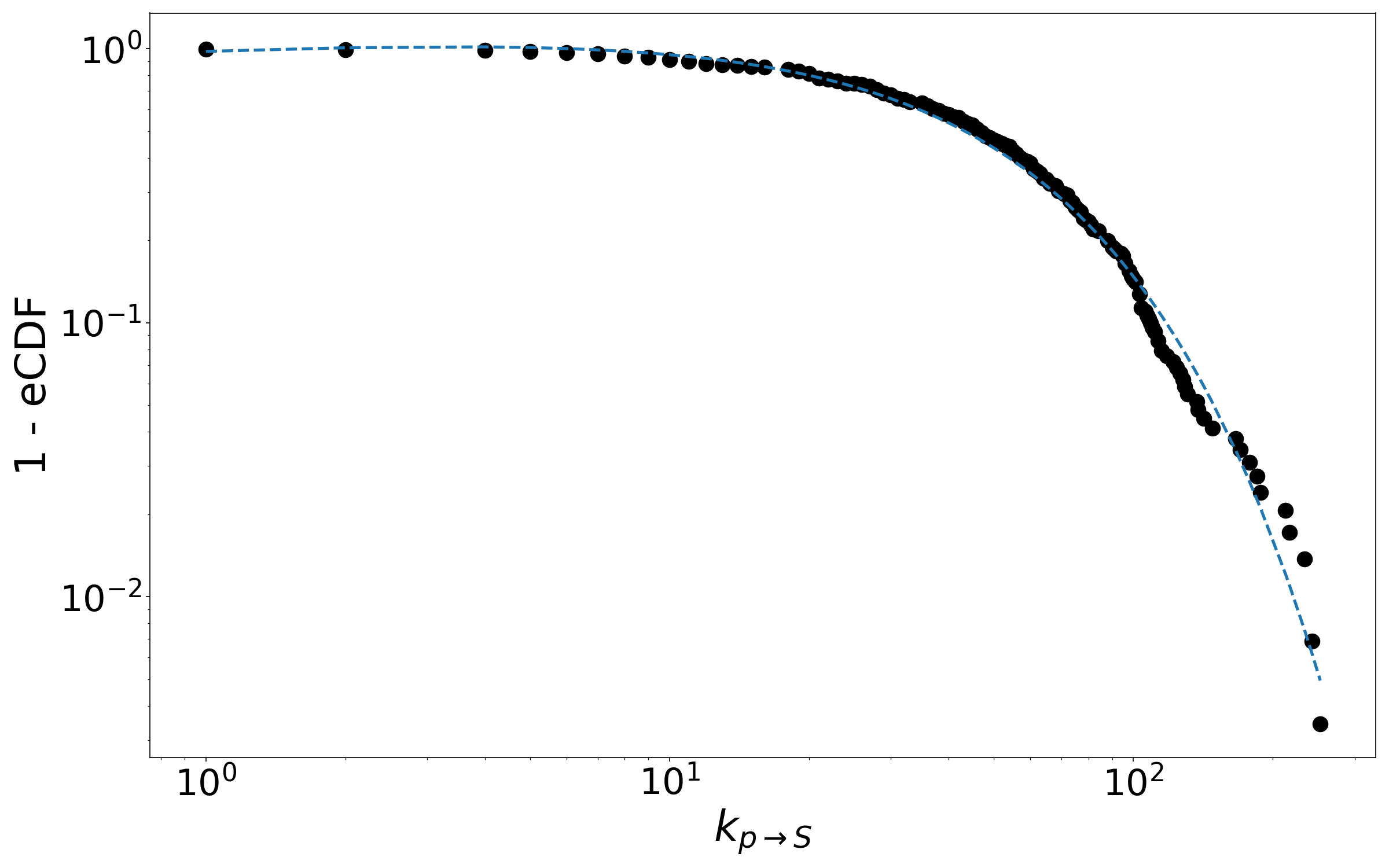}
\caption{Empirical cumulative density function of manufacturers' degrees (\textbf{A}), suppliers' `left' degrees (\textbf{B}), suppliers' `right' degrees (\textbf{C}) and products degrees (\textbf{D}). All of them are heavy-tailed and right-skewed, highlighting the large heterogeneity of the connectivity of nodes. More specifically, they all obey a power-law with exponential cutoff whose functional form reads $f(x)=x^{-\alpha}e^{-\beta x}\beta^{1-\alpha}/\Gamma[1-\alpha,\beta x_\text{min}]$ and whose parameters read $\alpha=0.134$, $\beta=0.005$ (top left), $\alpha=0.743$, $\beta=0.049$ (top right), $\alpha=1.276$, $\beta=0.015$ (bottom left), $\alpha=0.063$, $\beta=0.022$ (bottom right). These values have been determined by implementing the Levenberg-Marquardt algorithm~\cite {more2006levenberg} for least squares optimization through the Python library SciPy. Overall, the above results indicate that `generalist' suppliers co-exist with `specialist' suppliers - noticeably, $\simeq51\%$ of suppliers sells only one product.}
\label{fig2}
\end{figure*}

\paragraph*{Validating the monopartite projection.} P-values must, then, be validated by implementing a procedure for testing multiple hypotheses at a time. Several alternatives are viable, among which the Bonferroni correction \cite{tumminello2011statistically}, the Holm-Bonferroni correction and the Benjamini-Hochberg correction~\cite{thissen2002quick}. Here, we opt for the third one, controlling for the so-called False Discovery Rate (FDR), i.e. the expected proportion of false positives to appear within the set of tests that pass the validation. Thus, we sort the $n=M(M-1)/2$ p-values in increasing order

\begin{equation}
\text{p-value}_1\leq\text{p-value}_2\leq\dots\leq\text{p-value}_n
\end{equation}
and, then, individuate the largest integer $i$ satisfying the condition

\begin{equation}
\text{p-value}_i\leq\frac{it}{n}
\end{equation}
where $t$ represents the single-test significance level, set to the value of 0.01. The FDR procedure prescribes to reject the null hypothesis for all pairs of nodes whose p-value is less than, or equal to, $\text{p-value}_i$, meaning that their similarity is considered statistically significant - hence, not explainable by constraining (just) the degrees. The corresponding nodes are, thus, linked in the resulting, monopartite projection.\\

\subsection{Community detection}

In order to detect the presence of communities partitioning the node set of our monopartite, validated projections, we employ the popular, modularity-based Louvain algorithm~\cite{blondel2008fast}. Modularity is a score function that assesses the quality of a given partition by comparing the number of internal links with the one expected under a given null model: the Louvain algorithm implements a heuristic exploration of the landscape of partitions, individuating the one maximizing it.

Although faster and more accurate than other methods, the Louvain algorithm is sensitive to the order in which nodes are selected~\cite{lancichinetti2009community,fortunato2010community}. This limitation can be overcome by running it several times, each time considering nodes in a different order: the best partition will be, again, the one attaining the largest value of modularity.\\

To detect bipartite communities, instead, we employ the BiLouvain algorithm, an extension of the Louvain algorithm individuating the partition maximizing Barber's bipartite modularity~\cite{barber2007modularity}.

\section{Results and Discussion}

\begin{figure*}[t!]
\centering
\begin{minipage}{0.62\textwidth}
\textbf{A}\includegraphics[width=\textwidth]{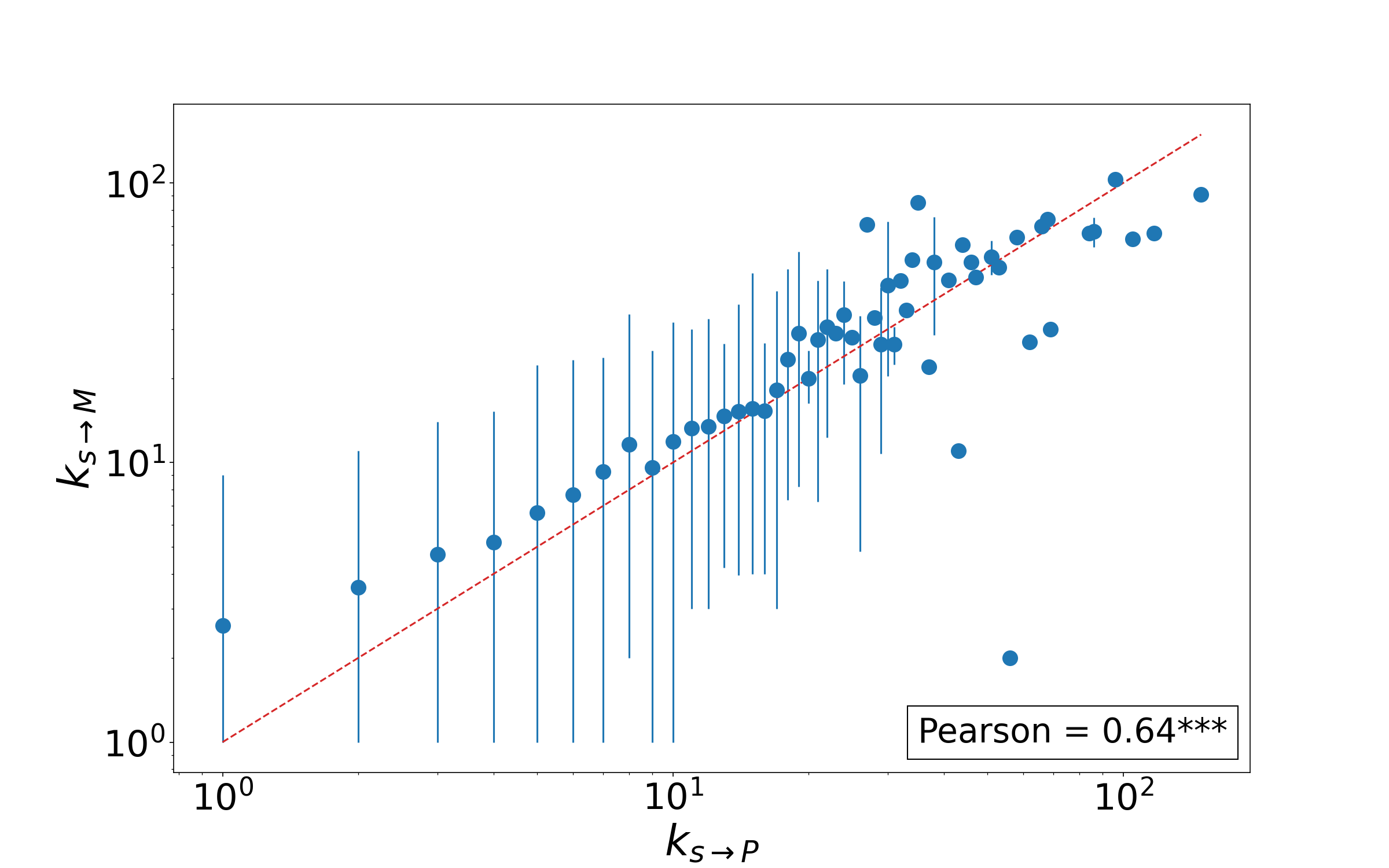}
\end{minipage}
\begin{minipage}{0.36\textwidth}
\textbf{B}\includegraphics[width=\textwidth]{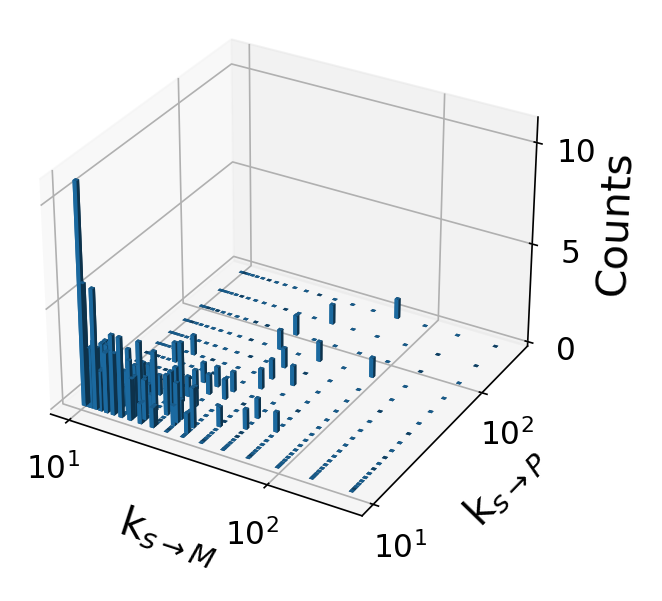}
\end{minipage}
\caption{\textbf{A}: average number of customers $k_{s\rightarrow\mathcal{M}}$ for suppliers with diversification $k_{s\rightarrow\mathcal{P}}$. The two quantities are positively correlated (the Pearson correlation coefficient reads $r\simeq0.64$ with a p-value smaller than 0.001), indicating that `generalist' suppliers tend to be connected with a large number of manufacturers; still, many suppliers providing few products while being connected with a relatively large number of manufacturers exist. The bars encode the $95\%$ of the sample distribution while the red, dashed line is the identity. \textbf{B}: 3D histogram of $k_{s\rightarrow\mathcal{M}}$ versus $k_{s\rightarrow\mathcal{P}}$ (displayed for the values $k_{s\rightarrow\mathcal{M}}>5$)}.
\label{fig3}
\end{figure*}

\begin{figure*}
\centering
\textbf{A}\includegraphics[width=0.48\textwidth]{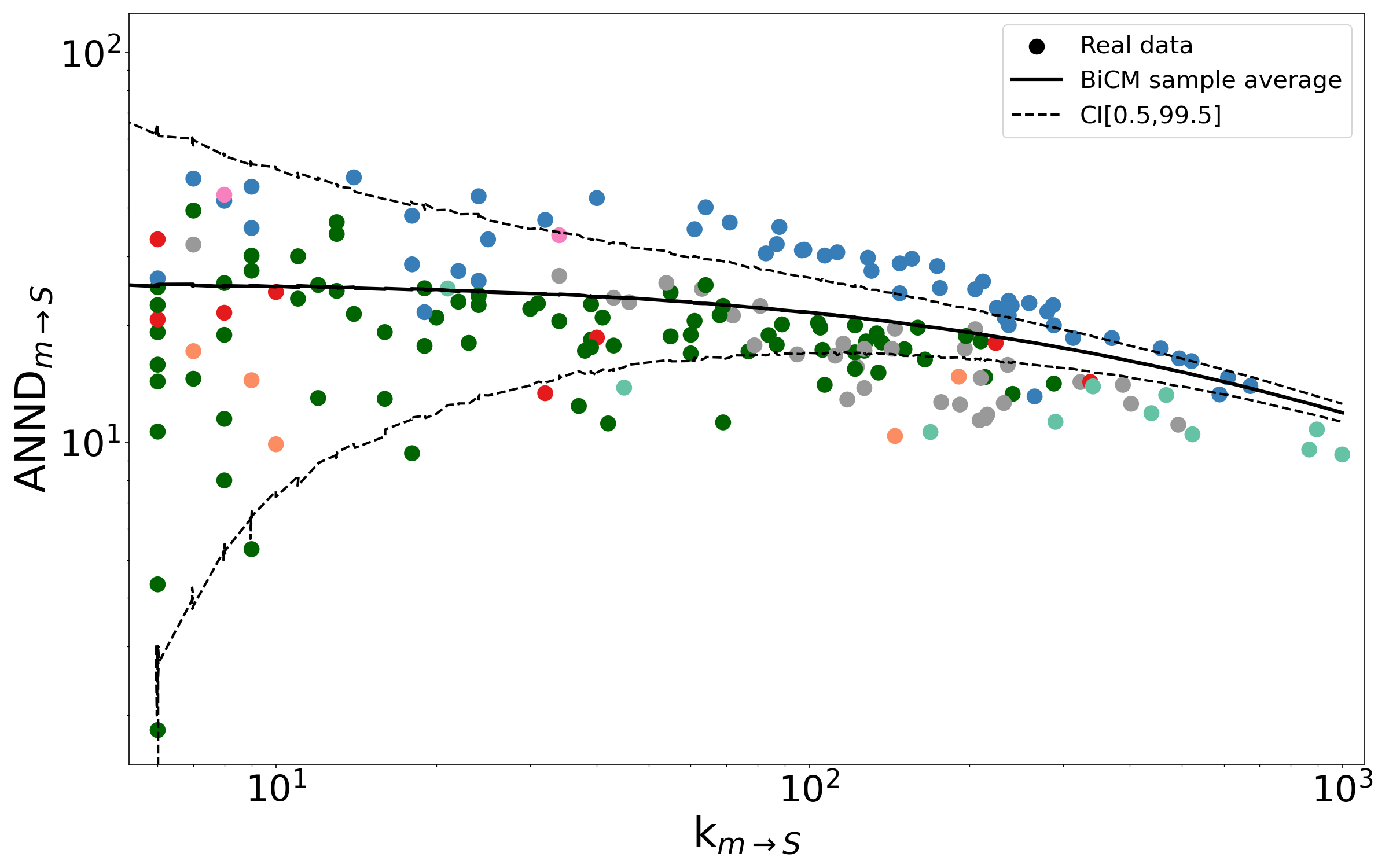}
\textbf{B}\includegraphics[width=0.48\textwidth]{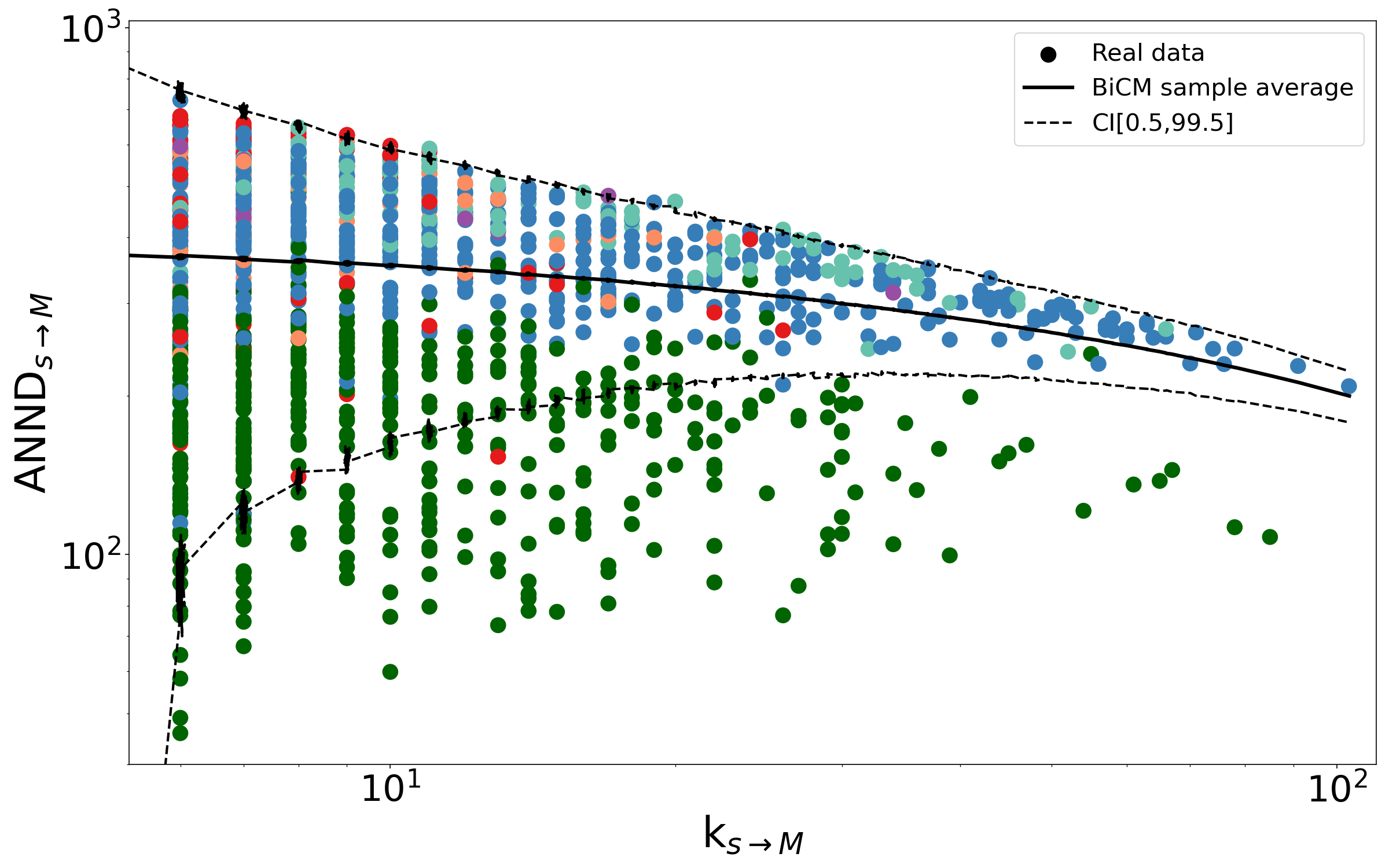}\\
\centering 
\textbf{C}\includegraphics[width=0.48\textwidth]{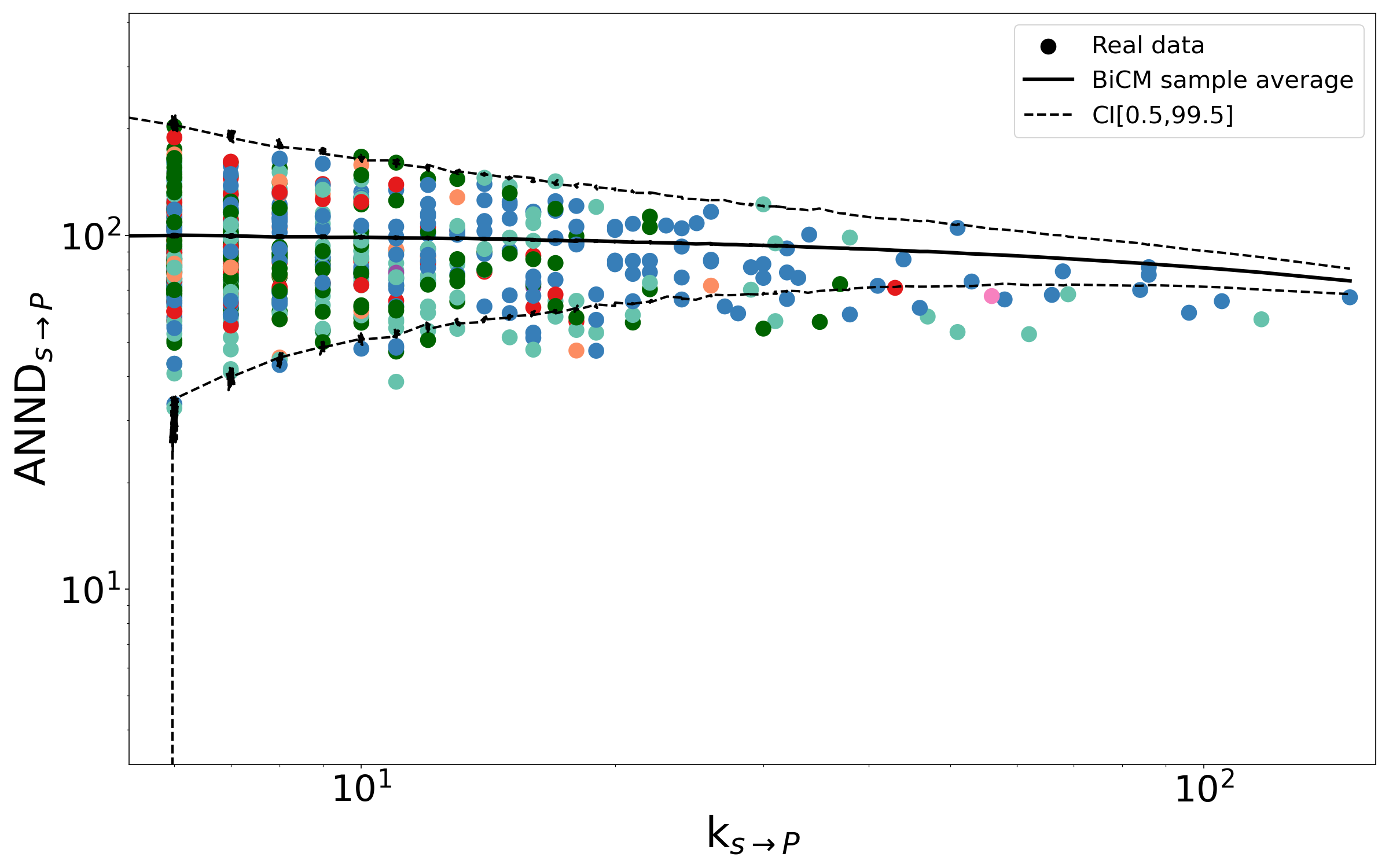}
\textbf{D}\includegraphics[width=0.48\textwidth]{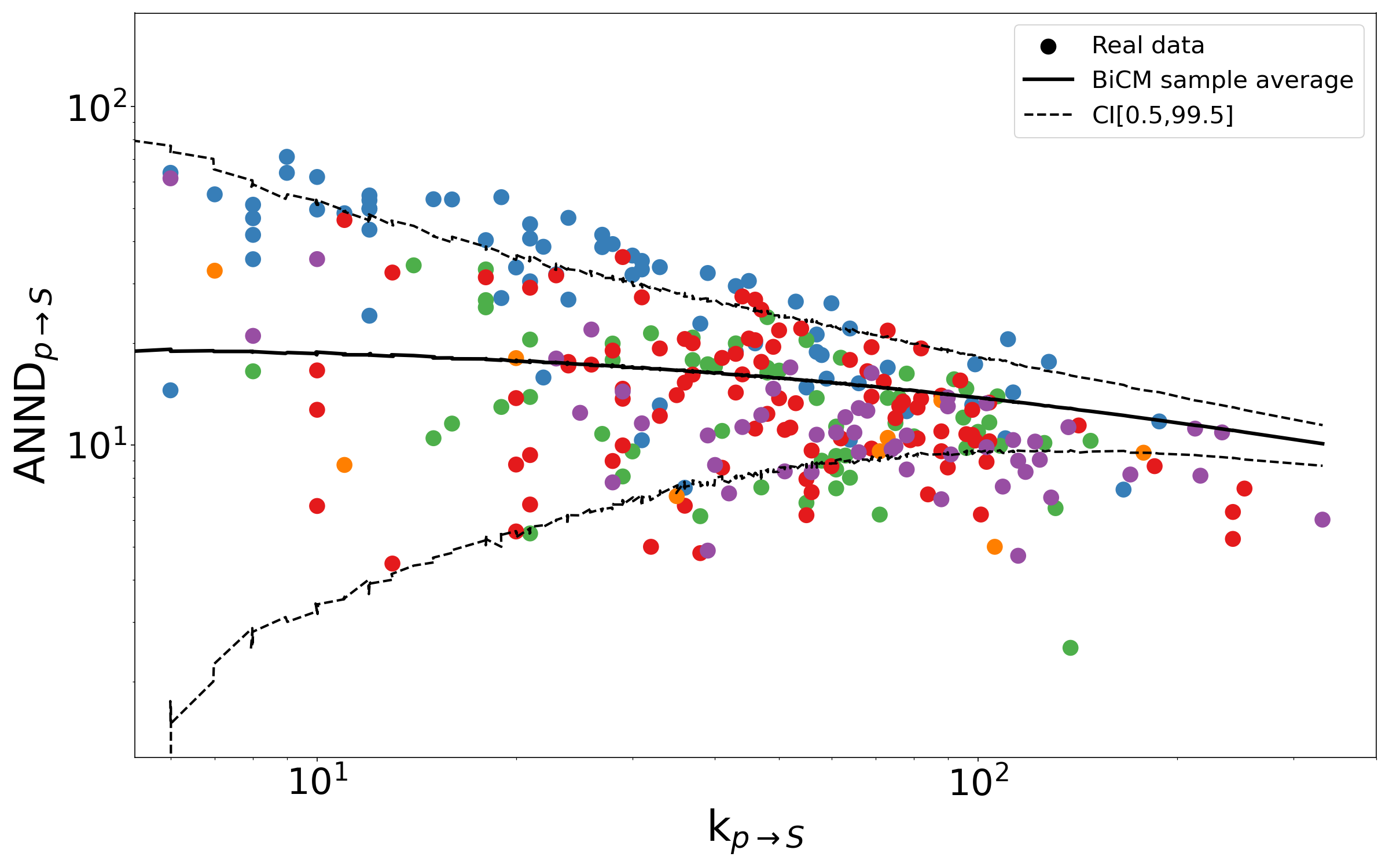}\\
\textbf{E}\includegraphics[width=0.48\textwidth]{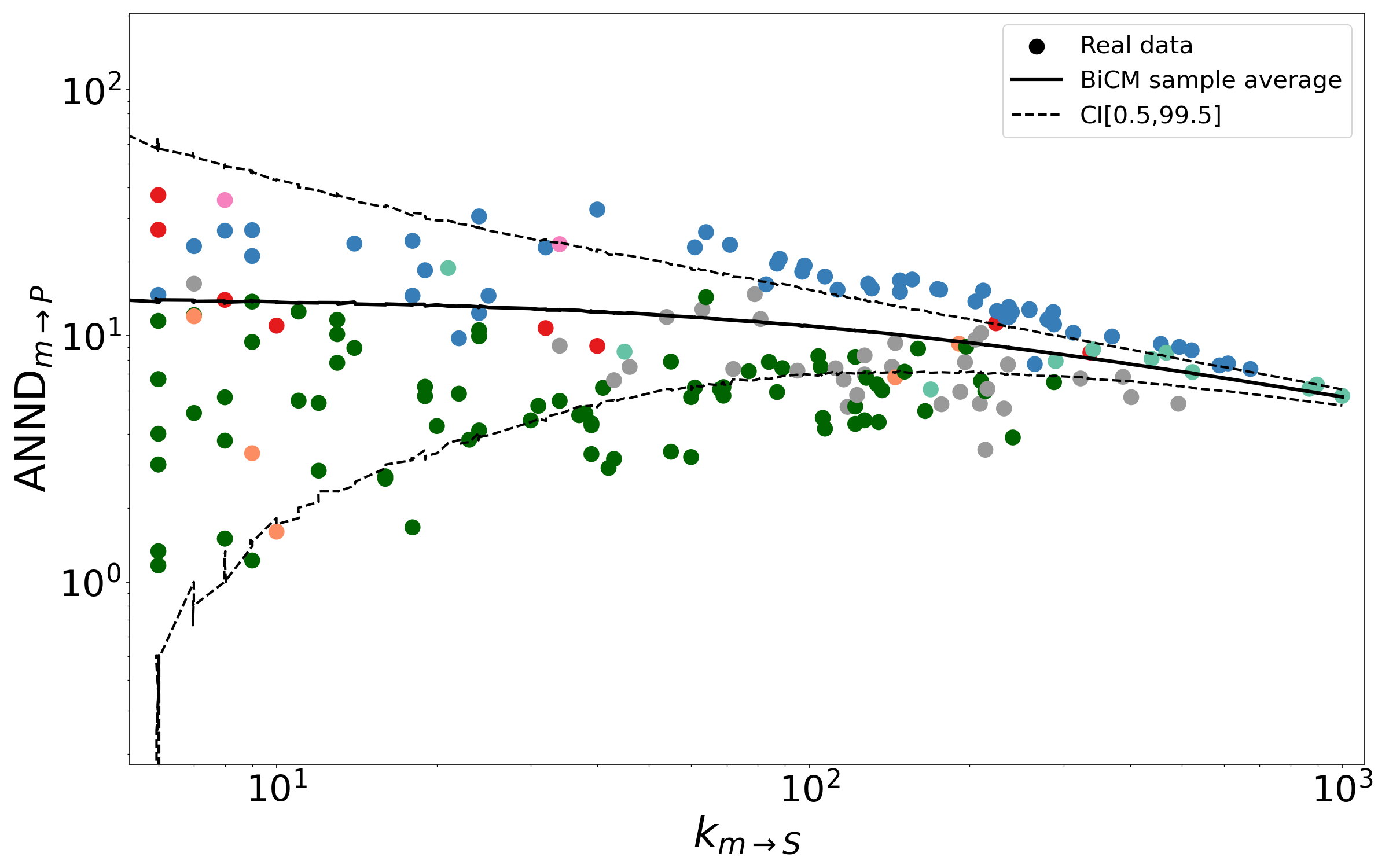}
\caption{Assortativity profiles of our networks. Top panels: $\text{ANND}_{m\rightarrow\mathcal{S}}$ scattered versus $k_{m\rightarrow\mathcal{S}}$ (\textbf{A}) and $\text{ANND}_{s\rightarrow\mathcal{M}}$ scattered versus $k_{s\rightarrow\mathcal{M}}$ (\textbf{B}). Both trends are slightly decreasing, hinting at a disassortative behavior, i.e. manufacturers (suppliers) having a large degree are generally connected with suppliers (manufacturers) having a small degree and viceversa. In other words, our results depict the automotive industry as an ecosystem populated by 1) suppliers that serve `bigger players' by providing them few products and 2) suppliers that serve a larger number of clients by providing them a larger basket of products. Middle panels: $\text{ANND}_{s\rightarrow\mathcal{P}}$ scattered versus $k_{s\rightarrow\mathcal{P}}$ (\textbf{C}) and $\text{ANND}_{p\rightarrow\mathcal{S}}$ scattered versus $k_{p\rightarrow\mathcal{S}}$ (\textbf{D}). As for the manufacturer-supplier network, both trends are decreasing, pointing out a disassortative behavior, according to which suppliers providing many products tend to have exclusive products within their baskets, whereas suppliers providing few products tend to have a basket of ubiquitous products (in this case, however, no geographically-induced distinction between firms emerges). \textbf{E}: $\text{ANND}_{m\rightarrow P}$ scattered versus $k_{m\rightarrow\mathcal{S}}$. The slightly decreasing trend confirms the picture provided by the figures above, i.e. Western and Japanese manufacturers with a larger degree tend to be connected with suppliers providing fewer products and viceversa. Nodes are colored according to their geographical localization: \textcolor{africa}{$\bullet$} Africa, \textcolor{asean}{$\bullet$} Asean, \textcolor{china}{$\bullet$} Chinese, \textcolor{india}{$\bullet$} Indian, \textcolor{japan}{$\bullet$} Japanese, \textcolor{southam}{$\bullet$} Latin America, \textcolor{middleeast}{$\bullet$} Middle East, \textcolor{russia}{$\bullet$} Russian, \textcolor{west}{$\bullet$} Western and \textcolor{jv}{$\bullet$} Joint Ventures. Products are colored according to their technological sector, i.e. \textcolor{cb}{$\bullet$} Chassis/Body, \textcolor{el}{$\bullet$} Electrical, \textcolor{pw}{$\bullet$} Powertrain, \textcolor{ie}{$\bullet$} Interior/Exterior, \textcolor{gp}{$\bullet$} General parts.}
\label{fig4}
\end{figure*}

\subsection{Structural properties}

\paragraph*{Local connectivity.} Figure~\ref{fig2} shows the degree distributions of manufacturers, suppliers and products. All of them are heavy-tailed and right-skewed, an evidence pointing out the large heterogeneity of nodes connectivity: more specifically, they all obey a power-law with exponential cutoff~\cite{clauset2009power}.\\

Overall, the above results indicate the presence of `generalist' suppliers (i.e. providing many products) co-existing with `specialist' suppliers (i.e. providing few products; $\simeq51\%$ of suppliers sells only one product). Figure~\ref{fig3} further shows that the number of client manufacturers of a supplier and the number of different products it sells are positively correlated, indicating that `generalist' suppliers tend to be connected with a large number of manufacturers. Still, many `specialist' suppliers that are linked to a relatively large number of manufacturers exist; a noticeable exception is \emph{Motor Super}, a Russian company that sells 56 different products to just 2 manufacturers, i.e. the Russian \emph{AvtoVaz} and the American \emph{Chevrolet}. Overall, $\simeq41\%$ of suppliers is connected to only one manufacturer.\\

\begin{figure*}[t!]
\centering
\textbf{A}\includegraphics[width=0.48\linewidth]{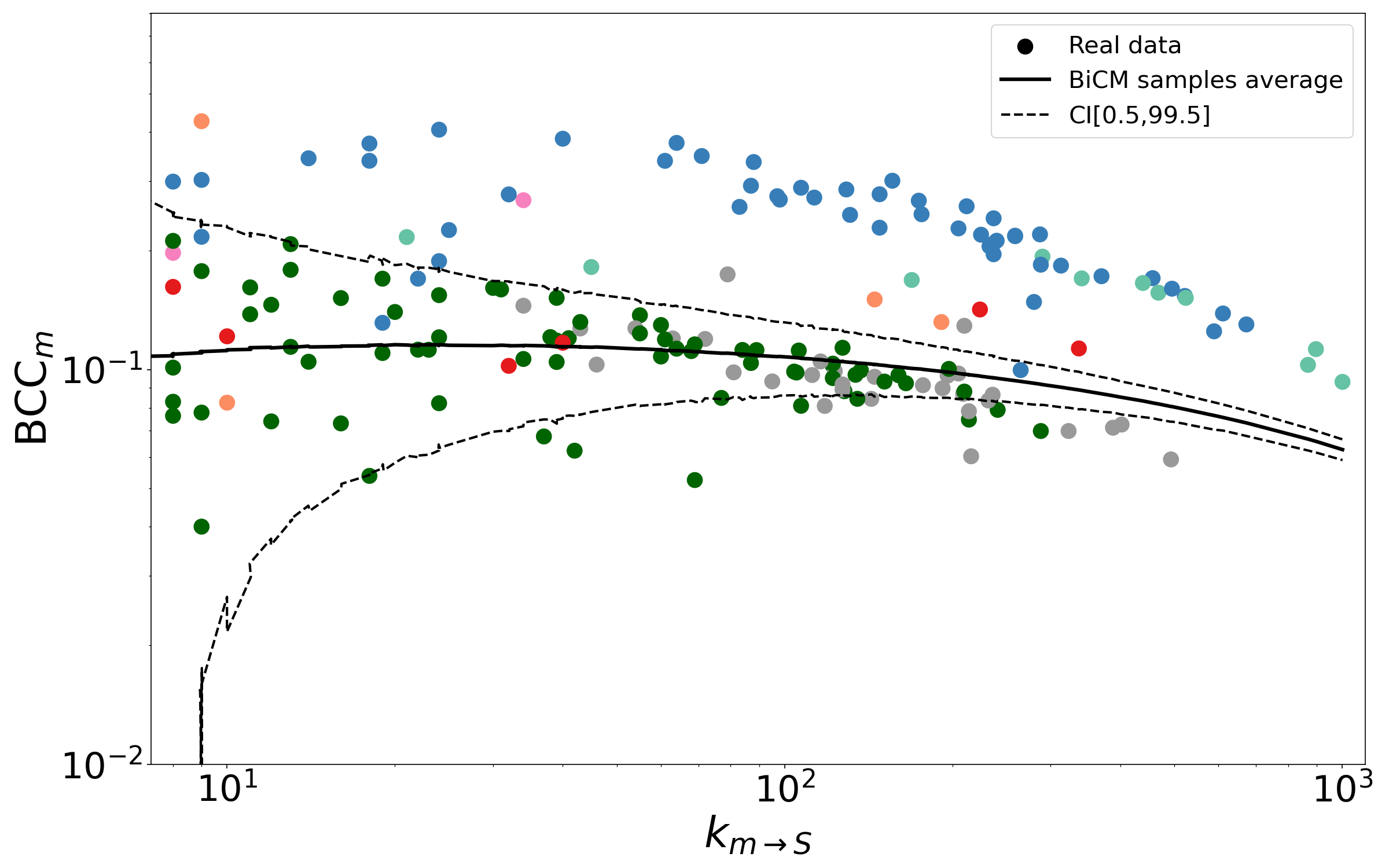}
\textbf{B}\includegraphics[width=0.48\linewidth]{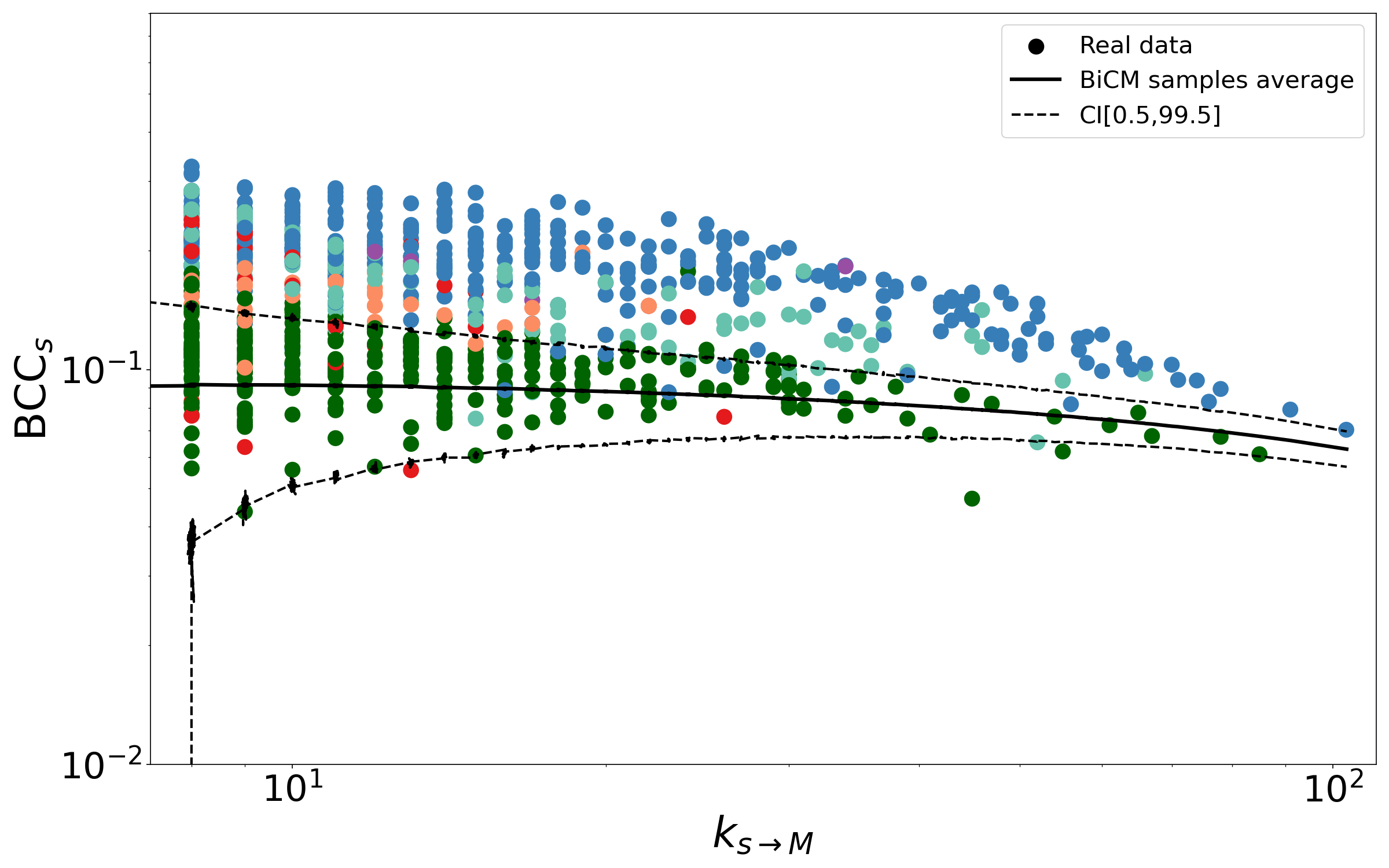}
\caption{Bipartite clustering coefficient of manufacturers $\text{BCC}_m$ scattered versus the number of their suppliers $k_{m\rightarrow\mathcal{S}}$ (\textbf{A}) and bipartite clustering coefficient of suppliers $\text{BCC}_s$ scattered versus the number of their client manufacturers $k_{s\rightarrow\mathcal{M}}$ (\textbf{B}). Both trends are overall decreasing, indicating that manufacturers (suppliers) having a large degree are generally closing less squares than manufacturers (suppliers) having a small degree. Besides, our results highlight that the automotive industry of different geographical locations is shaped by different organizing principles. The Western ecosystem seems to be highly interconnected, as manufacturers share many suppliers that, in turn, share many different manufacturers. Instead, the Chinese ecosystem seems to be rather fragmented, with different (sets of) manufacturers purchasing products from different (sets of) suppliers, each one serving few clients. Overall, our results confirm the presence of the (statistically significant) `functional' structures first observed in~\cite{mattsson2021functional}, however highlighting their peculiarly geographical character. Nodes are colored according to their geographical localization: \textcolor{africa}{$\bullet$} Africa, \textcolor{asean}{$\bullet$} Asean, \textcolor{china}{$\bullet$} Chinese, \textcolor{india}{$\bullet$} Indian, \textcolor{japan}{$\bullet$} Japanese, \textcolor{southam}{$\bullet$} Latin America, \textcolor{middleeast}{$\bullet$} Middle East, \textcolor{russia}{$\bullet$} Russian, \textcolor{west}{$\bullet$} Western and \textcolor{jv}{$\bullet$} Joint Ventures.}
\label{fig5}
\end{figure*}

\paragraph*{Assortativity.} In order to inspect the presence of degree correlations, we have scattered the $\text{ANND}_{m\rightarrow\mathcal{S}}$ values versus the $k_{m\rightarrow\mathcal{S}}$ values (Figure~\ref{fig4}\textbf{A}) and the $\text{ANND}_{s\rightarrow\mathcal{M}}$ values versus the $k_{s\rightarrow\mathcal{M}}$ values (Figure~\ref{fig4}\textbf{B}). As the plots show, both trends are slightly decreasing, pointing out the presence of disassortative patterns: in other words, manufacturers with many suppliers tend to be connected with suppliers having few customers, while manufacturers with few suppliers tend to be connected with suppliers having many customers; similarly, suppliers with many clients tend to be connected with manufacturers having few suppliers, while suppliers with few clients tend to be connected with manufacturers having many suppliers. Overall, then, the automotive industry resembles an ecosystem where suppliers serving (only) `bigger players', by providing them few products - the so-called `specialists' - co-exist with suppliers serving a larger number of clients, by providing them a larger basket of products - the so-called `generalists'.

To spot differences between firms induced by their geographical localization, we have partitioned them into nine groups, i.e. African, Asean (i.e. Indonesia, Malaysia, South Korea, Thailand, and Vietnam), Chinese, Indian, Japanese, Middle Eastern (i.e. Egypt, Iran, and Turkey), Russian, Western (i.e. Australia, Europe, Israel, and US) and Joint Ventures (JVs) and colored them accordingly. Overall, the group of Western firms is quite well-distinguished from the group of Chinese firms and JVs. Among the manufacturers, Western ones display significantly large $\text{ANND}_{m\rightarrow\mathcal{S}}$ values, lying in the top $0.5\%$ of the ensemble distribution induced by the null model, while the $\text{ANND}_{m\rightarrow\mathcal{S}}$ values for JVs, Chinese and Japanese manufacturers are either in line with the predictions of the null model or lie in the bottom $0.5\%$ of the null distribution. For what concerns suppliers, instead, Chinese firms display significantly small $\text{ANND}_{s\rightarrow\mathcal{M}}$ values, lying in the bottom $0.5\%$ of the null distribution while the $\text{ANND}_{s\rightarrow\mathcal{M}}$ values for Western and Japanese suppliers are in line with the predictions of the null model. This result suggests the Western and Japanese supply chains to be structurally different from the Chinese ones: Western and Japanese manufacturers tend to purchase products from suppliers whose degree is, on average, larger than the one of the suppliers serving Chinese manufacturers. In particular, the Japanese automotive industry revolves around few, big manufacturers purchasing products from many, low-degree suppliers.

For what concerns the supplier/product network, scattering the $\text{ANND}_{s\rightarrow\mathcal{P}}$ values versus the $k_{s\rightarrow\mathcal{P}}$ values and the $\text{ANND}_{p\rightarrow\mathcal{S}}$ values versus the $k_{p\rightarrow\mathcal{S}}$ values reveals its disassortative character, with suppliers selling many, less ubiquitous products and viceversa ((Figure~\ref{fig4}\textbf{C-D})). This result is reminiscent of the one concerning the export of countries within the (bipartite representation of the) international trade. While firms do not seem to be partitioned according to any geographical criterion, products belonging to the Electrical sector display the larger $\text{ANND}_{p\rightarrow S}$ values.

The aforementioned, geographical difference is recovered when scattering the tripartite assortativity coefficient of each manufacturer, defined as $\text{ANND}_{m\rightarrow\mathcal{P}}=\sum_{s=1}^Sl_{ms}k_{s\rightarrow\mathcal{P}}/k_{m\rightarrow\mathcal{S}}$, versus its degree. As Figure~\ref{fig4}\textbf{E} confirms, Western and Japanese manufacturers tend to connect with suppliers providing a number of products that is larger than the one provided by the suppliers to which Chinese manufacturers and JVs tend to connect.\\

\paragraph*{Motifs.} For what concerns the bipartite clustering coefficient, both panels of Figure \ref{fig5} show an overall decreasing trend, signaling that manufacturers (suppliers) having a large degree are generally closing less squares than manufacturers (suppliers) having a small degree. Again, the group of Western and Japanese firms is quite well-distinguished from the group of Chinese firms and JVs, as the former ones display significantly large values of the bipartite clustering coefficient, lying in the top $0.5\%$ of the ensemble distribution induced by our null model. In other words, the presence of mesoscale structures characterizing the Western and Japanese subsets of firms cannot be simply traced back to the degrees of nodes: rather, it represents a peculiar feature of these areas whose automotive industry appears as a highly interconnected ecosystem. Chinese firms, on the contrary, constitute a seemingly fragmented environment with different (sets of) manufacturers purchasing products from different (sets of) suppliers, each one serving few clients. Our results complement the ones about the presence of statistically significant functional structures within the automotive industry~\cite{mattsson2021functional}, clarifying that they come along a geographical signature. These results are robust with respect to the definition of the bipartite clustering coefficient (see Appendix B).\\

\begin{figure*}[t!]
\centering
\textbf{A}\includegraphics[width=.48\textwidth]{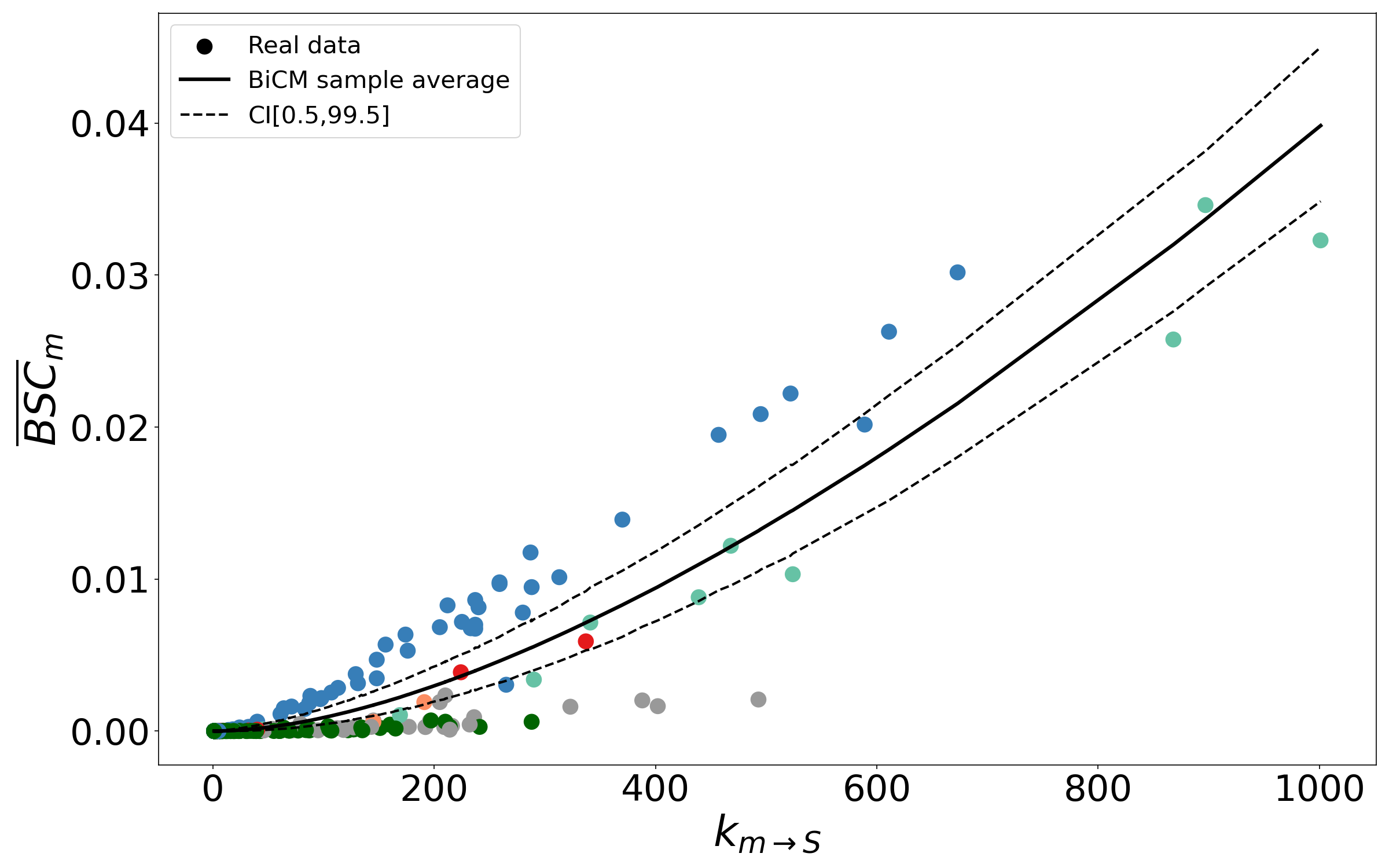}
\textbf{B}\includegraphics[width=.48\textwidth]{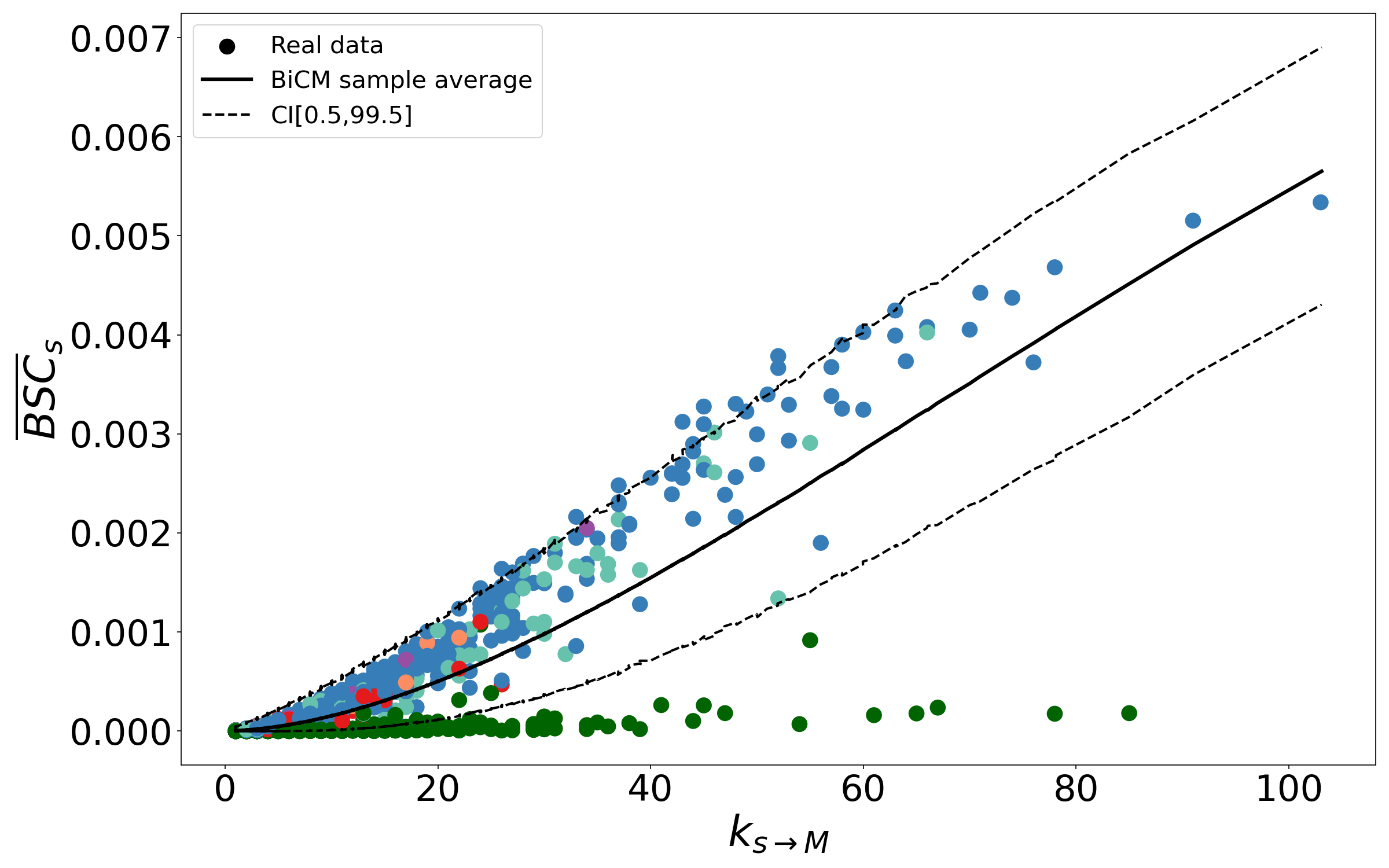}
\caption{Bipartite subgraph centrality of manufacturers $\overline{\text{BSC}}_m$ scattered versus the number of their suppliers $k_{m\rightarrow\mathcal{S}}$ (\textbf{A}) and bipartite subgraph centrality of suppliers $\overline{\text{BSC}}_s$ scattered versus the number of their client manufacturers $k_{s\rightarrow\mathcal{M}}$ (\textbf{B}). Both trends are increasing, pointing out that nodes with a larger degree are also more central. Chinese firms and JVs are characterized by values of the BSC that are much smaller than those characterizing Western and Japanese firms and always smaller-than-expected under the BiCM, a finding seemingly confirming that Chinese firms belong to less interconnected, if not completely segregated, supply chains. Western and Japanese firms, instead, appear to be crossed by a large number of patterns - actually, larger-than-expected under the BiCM, when considering the layer of manufacturers - an evidence confirming the plethora of interconnections leading from one node to another within this portion of the network. Nodes are colored according to their geographical localization: \textcolor{africa}{$\bullet$} Africa, \textcolor{asean}{$\bullet$} Asean, \textcolor{china}{$\bullet$} Chinese, \textcolor{india}{$\bullet$} Indian, \textcolor{japan}{$\bullet$} Japanese, \textcolor{southam}{$\bullet$} Latin America, \textcolor{middleeast}{$\bullet$} Middle East, \textcolor{russia}{$\bullet$} Russian, \textcolor{west}{$\bullet$} Western and \textcolor{jv}{$\bullet$} Joint Ventures.}
\label{fig6}
\end{figure*}

\paragraph*{Subgraph centrality.} The results of the analysis of the bipartite subgraph centrality (BSC), illustrated in Figure \ref{fig6}, show that nodes with a larger degree tend to be more central as well. More importantly, the BSC allows us to appreciate the different behavior displayed by firms located in different countries at best: Chinese firms and JVs are, in fact, characterized by values of the BSC that are much smaller (some hardly above zero) than the values of the BSC characterizing Western and Japanese firms. Specifically, the latter (former) have a larger-than-expected (smaller-than-expected) BSC on the layer of manufacturers. Similar trends are observed when considering the layer of suppliers, the only difference being that, now, the BSC of Western and Japanese firms is, overall, in line with the predictions of the BiCM. Once again, these findings reveal Western and Japanese firms to be crossed by a large number of patterns, an evidence confirming the plethora of interconnections leading from one node to another within this portion of the network. Chinese firms, instead, seemingly belong to less interconnected, if not completely segregated, supply chains.\\

\begin{figure*}[t!]
\centering
\includegraphics[width=.7\linewidth]{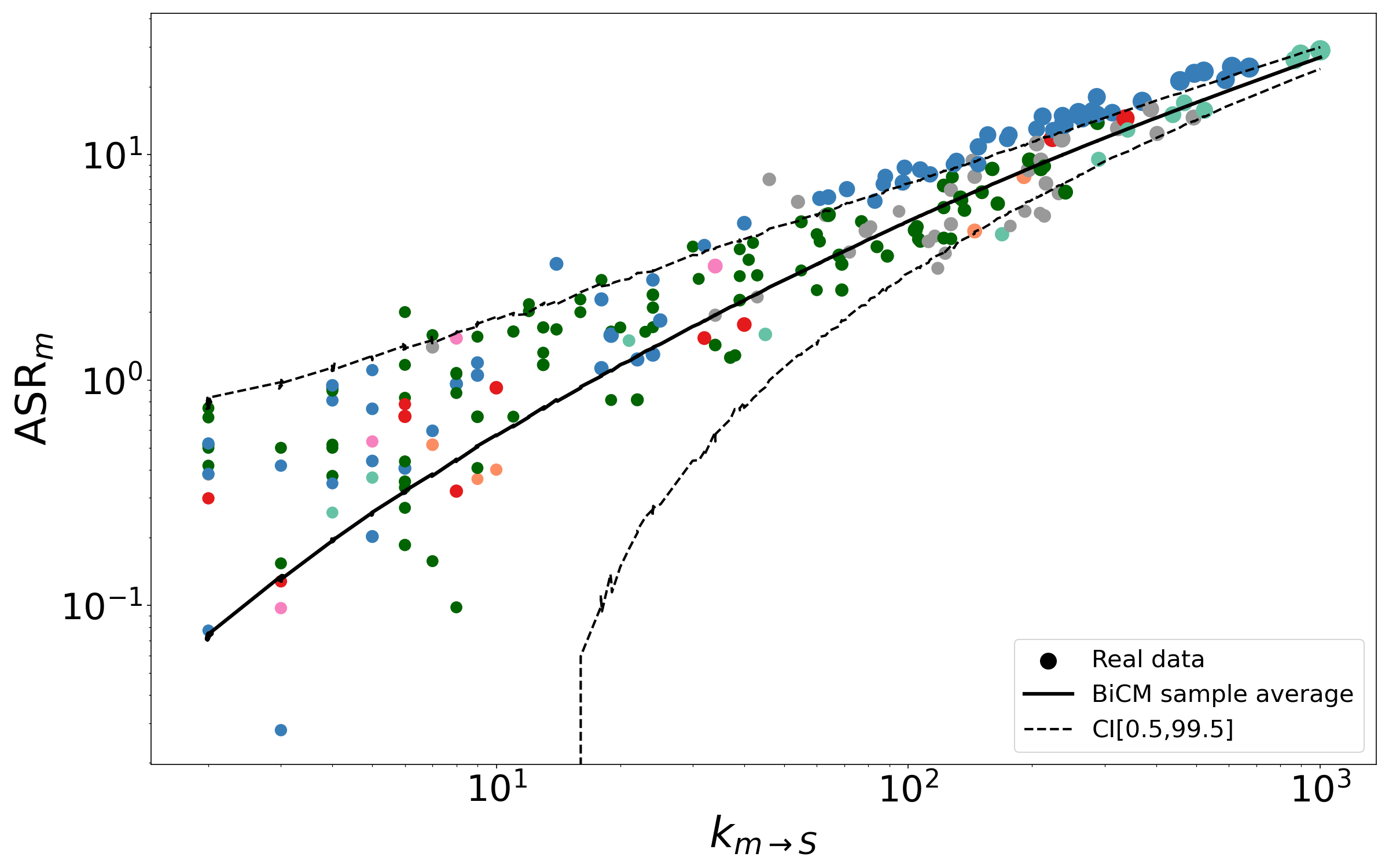}
\caption{Average suppliers redundancy of manufacturers $\text{ASR}_m$ scattered versus the number of their suppliers $k_{m\rightarrow S}$. The increasing trend signals that suppliers serving manufacturers with a larger degree tend to produce a larger number of similar products. The suppliers serving Western manufacturers display the larger redundancy values; the size of dots, proportional to the number of countries of the suppliers of each manufacturer, highlight a geographical signature: the suppliers serving Western and Japanese manufacturers are scattered across many, different countries (36 for \emph{Toyota} and 34 for \emph{Ford}) while this is true to a much lesser extent for what concerns the suppliers serving Chinese manufacturers and JVs (the number of countries hosting the suppliers of \emph{Geely} and \emph{FAW Volkswagen} are 10 and 15, respectively). Nodes are colored according to their geographical localization: \textcolor{africa}{$\bullet$} Africa, \textcolor{asean}{$\bullet$} Asean, \textcolor{china}{$\bullet$} Chinese, \textcolor{india}{$\bullet$} Indian, \textcolor{japan}{$\bullet$} Japanese, \textcolor{southam}{$\bullet$} Latin America, \textcolor{middleeast}{$\bullet$} Middle East, \textcolor{russia}{$\bullet$} Russian, \textcolor{west}{$\bullet$} Western and \textcolor{jv}{$\bullet$} Joint Ventures.}
\label{fig7}
\end{figure*}

\paragraph*{Suppliers redundancy.} We complement our analysis by calculating, for each manufacturer $m$, the average redundancy of its suppliers, defined as $\text{ASR}_m=\sum_{s=1}^Sl_{ms}\text{RED}_s/k_{m\rightarrow\mathcal{S}}$ where $\text{RED}_s=\sum_{p=1}^P\sum_{s'(\neq s)\in\mathcal{S}_m}r_{sp}r_{s'p}/k_{s\rightarrow\mathcal{P}}$ and with $\mathcal{S}_m$ indicating the neighborhood of manufacturer $m$, i.e. the set of its suppliers. Scattering the $\text{ASR}_m$ values versus the $k_{m\rightarrow\mathcal{S}}$ values (see Figure \ref{fig7}) reveals an increasing trend, pointing out that suppliers serving manufacturers with a larger degree tend to produce a larger number of similar products. Besides, the suppliers serving Western manufacturers display significantly large redundancy values, lying in the top $0.5\%$ of the ensemble distribution induced by our null model. The size of the dots is proportional to the geographical membership of the suppliers of each manufacturer: while the suppliers serving Western and Japanese manufacturers are scattered across many different countries - e.g. 36 for \emph{Toyota} and 34 for \emph{Ford} - this is true to a much lesser extent for Chinese manufacturers and JVs - e.g. the number of countries hosting the suppliers of \emph{Geely} and \emph{FAW Volkswagen} is 10 and 15, respectively.\\

\begin{table}[t!]
\caption{\label{tab2}General and in-block nestedness of the manufacturer/supplier and supplier/product networks, along with their expected values, standard deviations and $z$-scores, computed under the BiCM. While the general NODF is significantly smaller-than-expected, the opposite is true for its in-block version. Communities $l_{6}$ to $l_{10}$ and $r_{15}$ to $r_{19}$ are not shown as they are composed by just one or two nodes.}
\begin{ruledtabular}
\begin{tabular}{lcccc}
& \textbf{NODF} & $\langle$\textbf{NODF}$\rangle$ & $\mathbf{\sigma}_\text{NODF}$ & $z$-\textbf{score}\\
\hline
$\mathbf{L_{ms}}$ & \textbf{0.096} & \textbf{0.120} & \textbf{0.001} & \textbf{-14.542}\\
$\mathbf{l_{1}}$ & 0.119 & 0.077 & 0.002 & 17.941\\
$\mathbf{l_{2}}$ & 0.407 & 0.190 & 0.010 & 22.288\\
$\mathbf{l_{3}}$ & 0.424 & 0.244 & 0.007 & 24.462\\
$\mathbf{l_{4}}$ & 0.293 & 0.137 & 0.013 & 11.742\\
$\mathbf{l_{5}}$ & 0.219 & 0.076 & 0.017 & 8.484\\
\hline
$\mathbf{R_{sp}}$ & \textbf{0.023} & \textbf{0.031} & \textbf{0.0004} & \textbf{-19.859}\\
$\mathbf{r_{1}}$ & 0.086 & 0.073 & 0.005 & 2.599\\
$\mathbf{r_{2}}$ & 0.137 & 0.062 & 0.007 & 10.741\\
$\mathbf{r_{3}}$ & 0.078 & 0.036 & 0.010 & 4.314\\
$\mathbf{r_{4}}$ & 0.085 & 0.034 & 0.017 & 2.975\\
$\mathbf{r_{5}}$ & 0.124 & 0.049 & 0.012 & 6.084\\
$\mathbf{r_{6}}$ & 0.345 & 0.038 & 0.033 & 9.254\\
$\mathbf{r_{7}}$ & 0.308 & 0.068 & 0.050 & 4.788\\
$\mathbf{r_{8}}$ & 0.123 & 0.055 & 0.020 & 3.464\\
$\mathbf{r_{9}}$ & 0.112 & 0.041 & 0.015 & 4.848\\
$\mathbf{r_{10}}$ & 0.168 & 0.085 & 0.030 & 2.777\\
$\mathbf{r_{11}}$ & 0.130 & 0.026 & 0.024 & 4.244\\
$\mathbf{r_{12}}$ & 0.177 & 0.044 & 0.023 & 5.686\\
$\mathbf{r_{13}}$ & 0.396 & 0.082 & 0.035 & 8.960\\
$\mathbf{r_{14}}$ & 0.215 & 0.037 & 0.033 & 5.405\\
\end{tabular}
\end{ruledtabular}
\end{table}

\paragraph*{Nestedness.} Nestedness has been shown to characterize the `MarkLines Automotive' dataset~\cite{brintrup2012nested,brintrup2015nested,chauhan2021relationship}. Our analysis, however, returns a quite different picture, showing that $\mathbf{L}$ and $\mathbf{R}$ display smaller-than-expected values of the NODF (see Table \ref{tab2}). Left aside the differences concerning the data (the authors of~\cite{brintrup2015nested} consider 2.474 manufacturers, 16.468 suppliers and 934 products), we argue this discrepancy to be traceable back to 1) the way nestedness is computed (namely, by picking random subsamples composed of 50 rows~\cite{brintrup2015nested}), 2) the different randomization strategy implemented.

In order to further inspect the mesoscopic structure of our networks, we looked for the presence of nested communities. To this aim, we, first, performed a bipartite community detection on $\mathbf{L}$ and $\mathbf{R}$ and, then, measured the in-block nestedness. Both the manufacturer/supplier and the supplier/product networks display a marked, bipartite community structure, with bi-modularity amounting at $Q=0.51$ and $Q=0.57$, respectively. The former contains 10 communities, the five, biggest ones clearly identifying the Asean-Indian, Chinese, European, Japanese and US clusters (see Figure \ref{fig8}); the latter contains 19 communities, representing coherent sets of products. In both cases, the in-block NODF is characterized by larger-than-expected values (see Table \ref{tab2}). If, instead, we test nestedness against the block-specific version of the BiCM, we recover smaller-than-expected values.

\subsection{Projections of the `MarkLines Automotive' production network}

\begin{figure*}[t!]
\centering
\textbf{A}\includegraphics[width=0.7\linewidth]{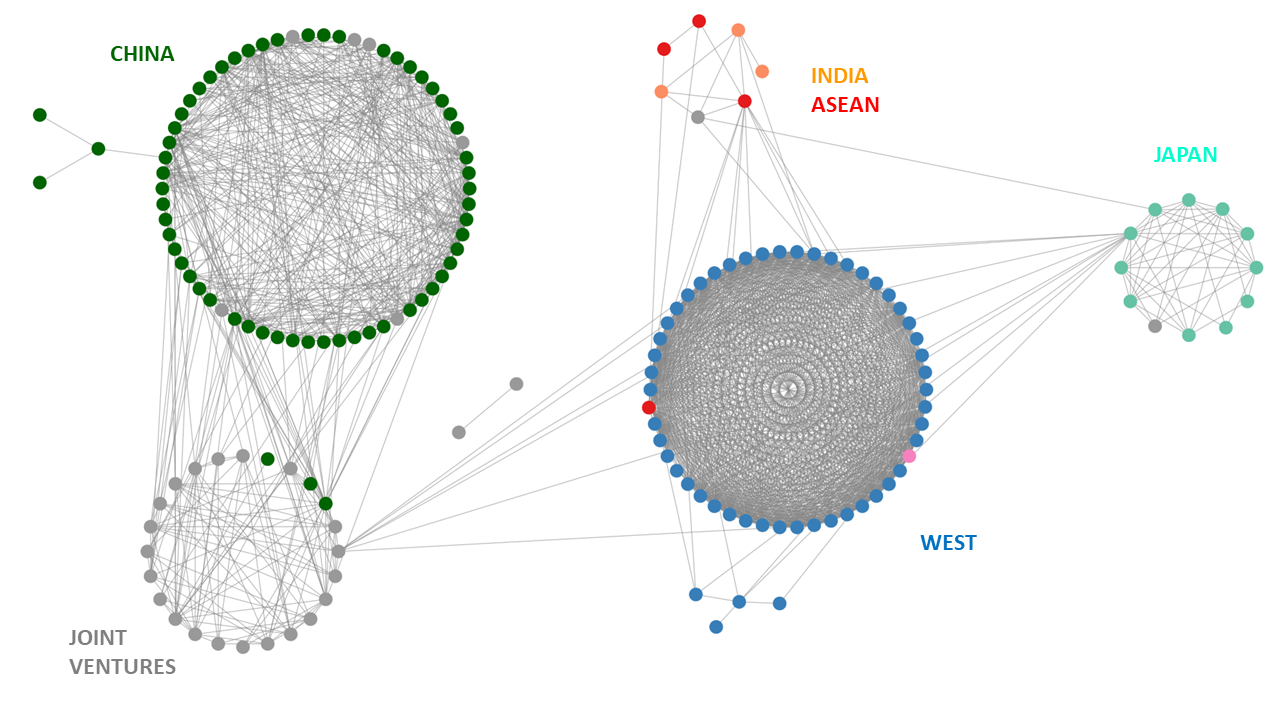}
\textbf{B}\includegraphics[width=0.685\linewidth]{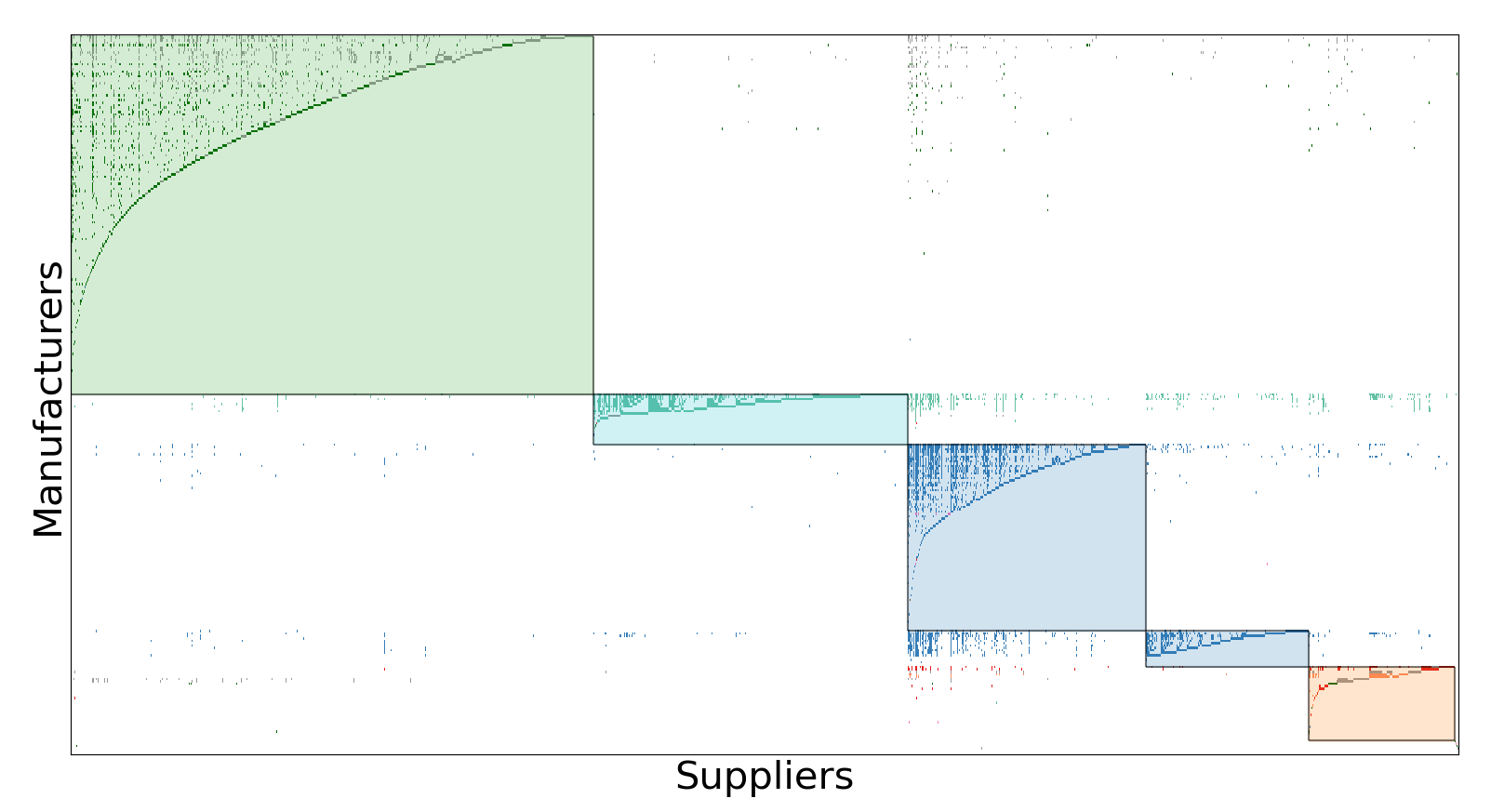}
\caption{\textbf{A}: validated projection onto the layer of manufacturers, linked if sharing a significantly large number of suppliers. Analyzing the presence of connections between communities reveals that the Chinese cluster is solely linked to the cluster of JVs which, in turn, is connected with the Western cluster via the manufacturer \emph{BMW Brilliance}. As the latter is connected with the Asean-Indian and Japanese clusters as well, it is also the most central one of our validated projection. Lastly, the Western cluster is much more internally connected than the Chinese one, a finding further confirming the closure of motifs to be strongly driven by geographical proximity. \textbf{B}: biadjacency matrix of the manufacturer/supplier network, with nodes partitioned according to the communities spotted by the BiLouvain algorithm and reordered according to the `Economic Fitness and Complexity' algorithm~\cite{tacchella2012new}. The triangularity of each module confirms the in-block nested structure of the network. Notice the overlap between the two sets of communities.} Nodes are colored according to their geographical localization: \textcolor{africa}{$\bullet$} Africa, \textcolor{asean}{$\bullet$} Asean, \textcolor{china}{$\bullet$} Chinese, \textcolor{india}{$\bullet$} Indian, \textcolor{japan}{$\bullet$} Japanese, \textcolor{southam}{$\bullet$} Latin America, \textcolor{middleeast}{$\bullet$} Middle East, \textcolor{russia}{$\bullet$} Russian, \textcolor{west}{$\bullet$} Western and \textcolor{jv}{$\bullet$} Joint Ventures.
\label{fig8}
\end{figure*}

\paragraph*{Community structure of the validated projections.} Lastly, we focus on the validated projection onto each of our layers. Figure \ref{fig8}A shows the validated network of manufacturers, where any two nodes are linked if sharing a significantly large number of suppliers. A markedly modular structure emerges (the value of modularity amounts to $Q=0.50$), with clusters of manufacturers that are coherent with their geographical localization. Indeed the two, largest ones are those composed by Chinese and Western firms - which, however, are not interconnected. The Chinese cluster, instead, is linked only to the community of JVs, constituted by Joint Ventures involving one Chinese company. In turn, the Western cluster is connected with the cluster of JVs via the manufacturer \emph{BMW Brilliance}, with the Japanese cluster via the manufacturer \emph{Nissan} and with the Asean-Indian cluster via several links. Interestingly, the similarity of the two, small sets of nodes (recognized by the Louvain algorithm as individual communities) which are detached from the Chinese and Western clusters is not only due to geographical proximity but also to their technological characterization. Indeed, the three nodes lying on the left of the Chinese cluster (i.e. \emph{GAC Aion}, \emph{Weltmeister} and \emph{Xpeng}) are all manufacturers of electric cars, while the four nodes lying below the Western cluster (i.e. \emph{Alpina}, \emph{Lotus}, \emph{McLaren} and \emph{SRT}) are all manufacturers of sportive cars. We stress once more that the Western cluster is much more internally-connected than the Chinese one, a finding further confirming that the closure of motifs, within the automotive industry, is strongly driven by geographical proximity.

\begin{figure*}[t!]
\centering
\includegraphics[width=\linewidth]{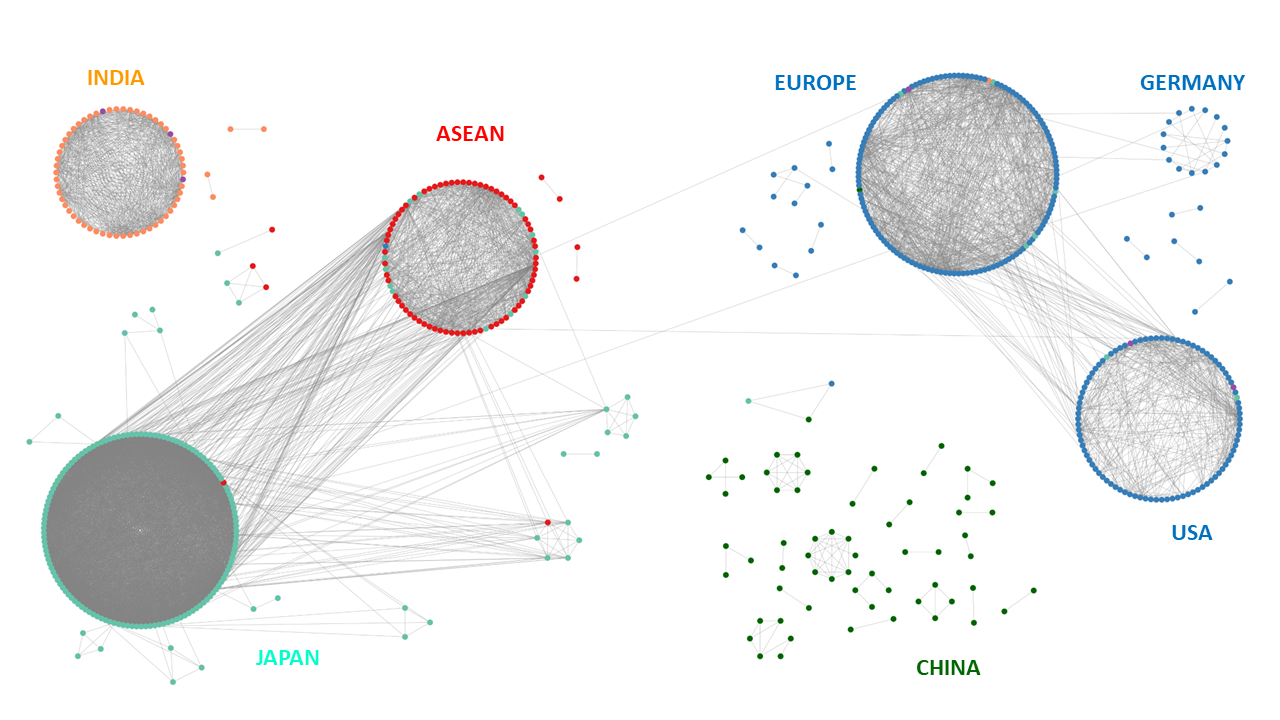}
\caption{Validated projection of suppliers, linked if sharing a significantly large number of manufacturers. Analyzing the presence of connections between communities reveals that the Chinese cluster is isolated and internally fragmented. The largest components are constituted by two pairs of interconnected communities, formed by Asean and Japanese firms and American and European firms. Nodes are colored according to their geographical localization: \textcolor{africa}{$\bullet$} Africa, \textcolor{asean}{$\bullet$} Asean, \textcolor{china}{$\bullet$} Chinese, \textcolor{india}{$\bullet$} Indian, \textcolor{japan}{$\bullet$} Japanese, \textcolor{southam}{$\bullet$} Latin America, \textcolor{middleeast}{$\bullet$} Middle East, \textcolor{russia}{$\bullet$} Russian, \textcolor{west}{$\bullet$} Western and \textcolor{jv}{$\bullet$} Joint Ventures.
}
\label{fig9}
\end{figure*}

Figure \ref{fig9} shows the validated projection of suppliers, where any two nodes are linked if sharing a significantly large number of manufacturers (see Appendix C). The network is markedly modular as well (the value of modularity amounts to $Q=0.55$) with clusters embodying geographical information. As evident upon inspecting the figure, the Chinese cluster is not only isolated but also (internally) fragmented, being constituted by a plethora of smaller connected components. The Indian cluster is disconnected from the rest of the network as well, although its density is quite large. The largest components are constituted by two pairs of interconnected communities, i.e. the one gathering Asean and Japanese firms and the one gathering American and European firms. German suppliers give origin to a smaller community, lying on the right of the European subgraph.

\begin{figure*}[t!]
\centering 
\includegraphics[width=\textwidth]{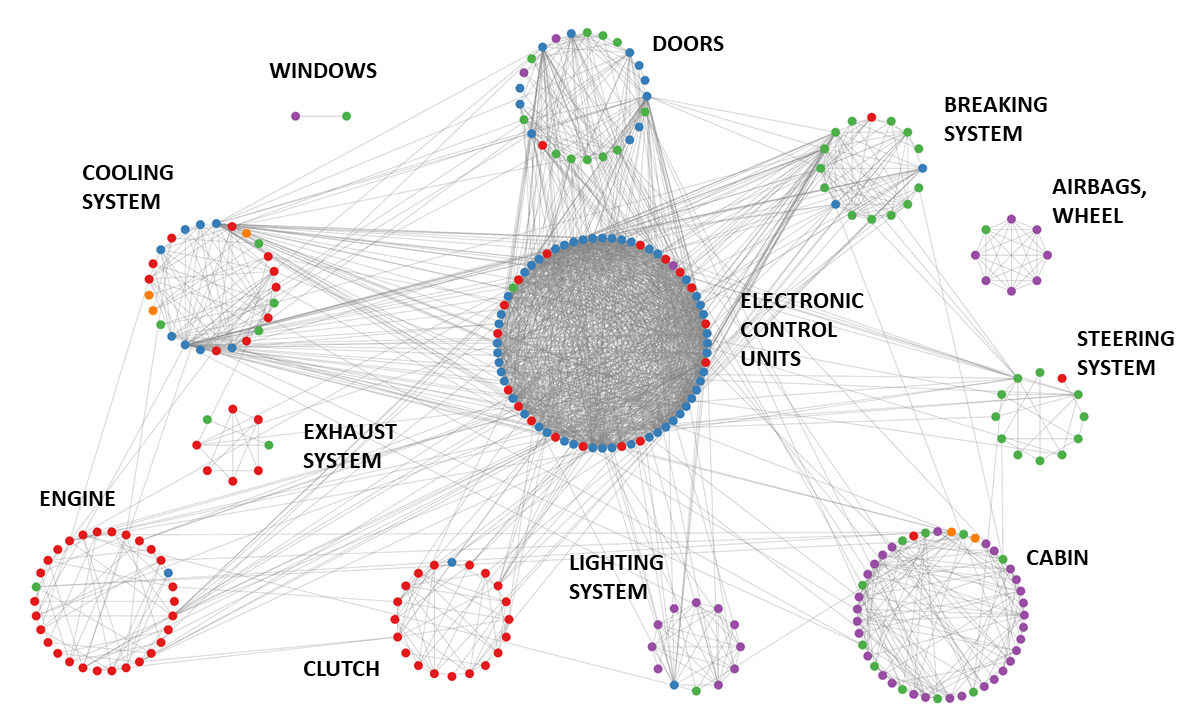}
\caption{Validated projection of products, linked if sharing a significantly large number of suppliers. The network displays a clearly defined community structure whose clusters represent different car systems. Nodes are colored according to their technological sector, i.e. \textcolor{cb}{$\bullet$} Chassis/Body, \textcolor{el}{$\bullet$} Electrical, \textcolor{pw}{$\bullet$} Powertrain, \textcolor{ie}{$\bullet$} Interior/Exterior, \textcolor{gp}{$\bullet$} General parts.
}
\label{fig10}
\end{figure*}

For what concerns the supplier/product network, the validated projection of products (linked if sharing a significantly large number of suppliers) is shown in Figure \ref{fig10}: it displays an interesting community structure (the value of modularity amounts to $Q=0.45$), characterized by the emergence of clusters that do not overlap with the (technological) taxonomy provided by the MarkLines platform. Rather, they represent different car systems (i.e. groups of devices targeting specific tasks - see Appendix D), an evidence suggesting that suppliers tend to focus their production on `coherent' sets of products~\cite{laudati2023different,albora2022machine}. Inspecting the nationality of the suppliers providing these products (see Appendix E) reveals that China, Germany, Japan and USA are the most represented countries. China, however, is extremely active in some communities (e.g. \emph{Airbags}, \emph{Engine}, \emph{Exhaust System}, \emph{Wheel}) while being completely absent from others (e.g. \emph{Electronic Control Units}); the percentage of suppliers of Germany, Japan and USA, instead, is similar across all communities.

\section{Conclusions}

The recent pandemic has fostered research on the structure of global supply chains~\cite{guan2020global,chowdhury2021covid}. The use of tools routinely employed in economics~\cite{taglioni2016making} has allowed scholars to approach a number of relevant, economic and geopolitical issues, such as the consequences of different, national development strategies on the global value chain of low-carbon technologies~\cite{goldthau2020protect} and the effects of the US-China trade war~\cite{fajgelbaum2021us}. Yet, these discussions have been typically carried out from a qualitative, or merely statistical, point of view. Additionally, commonly use databases such as input-output tables~\cite{timmer2015illustrated} are unable to resolve interactions at firm-level, since focusing on whole sectors in specific, geographical areas. The aim of the present paper is bridging this gap, carrying out a quantitative investigation of the large-scale structure of a specific industry by employing methods rooted into statistical physics. Our analysis of the `MarkLines Automotive' dataset allows us to draw a number of conclusions.

\begin{itemize}
\item As we obtain network projections by linking any two nodes in case the number of their common neighbors is significantly large and such a number is proportional to the number of X-motifs, we confirm the prominent role played by square patterns in shaping the structure of production networks~\cite{mattsson2021functional}: firms with similar outputs are often engaged in a competition, rather than in a buyer-supplier relationship, although they can share many suppliers and customers.

\item Degree distributions are heavy-tailed, hence describing an ecosystem where nodes are characterized by a highly heterogeneous number of neighbors. More explicitly, `generalist' suppliers selling many products co-exist together with `specialist' suppliers selling few products. Besides, `generalist' suppliers tend to be connected with a large number of small-degree manufacturers while `specialist' suppliers tend to be connected with a small number of large-degree manufacturers. Specifically, the suppliers selling only one product (amounting to $\simeq51\%$ of the total) are connected with a subset of 166 manufacturers whose average degree amounts to $\simeq151$, i.e. almost twice the manufacturers average degree - a result indicating that well-connected manufacturers are preferentially connected with suppliers seemingly providing `exclusive' services.

\item The analysis of the bipartite subgraph centrality points out that both manufacturers and suppliers with a larger degree tend to be more central, although the abundance of squares shows a decreasing trend when plotted as a function of the degree. These two quantities provide the neatest, geography-based partition of our basket of firms, separating Chinese from Western ones: the former close a smaller number of squares than the latter, hence providing an indication that the Chinese ecosystem is less integrated than the Western one. This is particularly evident when considering the set of Chinese suppliers: the limited, although statistically significant, number of neighbours they share induces a projection that is made of several disconnected components. Although this may be (at least, partly) induced by the peculiar way this projection has been obtained, our conclusions are supported by the (methodologically different) analysis carried out in~\cite{veloso2002automotive}, which finds that the Chinese automotive industry is characterized by a large degree of internal fragmentation - in play even at a territorial level, as the recovered components correspond to different, Chinese provinces. This evidence suggests the Chinese industrial organization to be more distributed than the Western one, maybe to ensure all provinces a comparable level of economic development.

\item The information provided by redundancy complements the one provided by geography, clarifying that the presence of (many) suppliers providing the same products to a manufacturer is compatible with a broader geographical distribution of the manufacturers plants. By converse, the presence of (fewer) suppliers selling different products to a manufacturer is compatible with a narrower geographical distribution of the manufacturers plants - in certain cases, a very local one. This result also suggests the manufacturers belonging to the first group are more resilient than the ones belonging to the second group, as their production is apparently less prone to interruptions due to supply shortages.

\item Projecting on the layer of manufacturers reveals that the Chinese cluster is solely connected with the cluster of Chinese JVs, a result indicating that Chinese manufacturers do not share (a significantly large number of) suppliers with other manufacturers. Again, this is supported by the analysis carried out in~\cite{veloso2002automotive}, where it has been observed that the Chinese government adopted protectionist policies to boost the development of an indigenous automotive industry, forcing Chinese JVs to buy the $40\%-80\%$ of their components from Chinese suppliers.
\end{itemize}

Of course, our study comes with limitations: we have only considered the data of a single industry (which may be incomplete as a result of the data collection procedure); our dataset has been represented in a bipartite fashion while, in general, production networks have a less neat structure; we have employed a null model embodying the information encoded into the node degrees although more refined benchmarks can be used; we have employed the Louvain algorithm which is one out of many alternative choices for the task of detecting communities (however, known to perform satisfactorily in presence of few, large communities).

As a last remark, we would like to stress that the investigation of the economic, historical, and social motivations at the origin of the geography-related structures we found is beyond the scope of the present paper. Here, we have shown that adopting tools from network theory can lead to the discovery of clear, structural features (in this case, concerning the automotive sector) that should not be overlooked during the model-building phase. These patterns, in fact, could inform methods for reconstructing production networks~\cite{cimini_mastrandrea_squartini_2021} - which aim at overcoming the limitations affecting available datasets by adopting statistically-grounded techniques~\cite{ialongo2022reconstructing}, machine learning tools~\cite{brintrup2018predicting,wichmann2020extracting,kosasih2021machine,Mungo:2023aa} or proper data proxies~\cite{reisch2022monitoring} - to be later employed for stress testing~\cite{fujiwara2016debtrank,shao2018data,diem2022quantifying}.

\section*{Acknowledgements}

This work is supported by the following projects: `SoBigData.it - Strengthening the Italian RI for Social Mining and Big Data Analytics', NextGenerationEU PNNR Grant IR0000013; `Network analysis of economic and financial resilience', Italian DM n. 289, 25-03-2021 (PRO3 Scuole) CUP D67G22000130001; 
`RENet - Reconstructing economic networks: from physics to machine learning and back', MUR PRIN 2022MTBB22 funded by European Union – Next Generation EU; `C2T - From Crises to Theory: towards a science of resilience and recovery for economic and financial systems', MUR PRIN PNRR P2022E93B8 funded by European Union – Next Generation EU; `WECARE - WEaving Complexity And the gReen Economy', MUR PRIN 20223W2JKJ funded by European Union – Next Generation EU.

\bibliography{main}

\clearpage

\section{Appendix A.\\Data cleaning and harmonization}

Here we detail the procedure used to clean and harmonize the `MarkLines Automotive' dataset.\\

First, we dealt with the presence of companies, reported along with their divisions, with multiple names that do not necessarily correspond to the actual taxonomy of the firm (e.g. `BAIC Motor' appears as `BAIC', `BAIC Motor' and `BAIC Group Off-road Vehicles'). Name homogenization was carried out by reconstructing the actual taxonomy of each group. Specifically, the corrections introduced are listed below:

\begin{itemize}
\item the instances of `BAIC' with \emph{Senova} and \emph{Beijing} models, `BAIC Motor' and `BAIC Group Off-road Vehicles' were listed as `BAIC Motor';
\item the instances of `BAIC' with \emph{Weiwang} models were listed as `BAIC Yinxiang';
\item `Changan', `Chongqing Changan' and `Changan Commercial Vehicles' were listed as `Chongqing Changan';
\item `Dongfeng', `Dongfeng Motor' and `Dongfeng Passenger Vehicles' were listed as `Dongfeng Motor';
\item the instances of `FAW', `FAW Car' with \emph{Bestune} models and `FAW Bestune' were listed as `FAW Bestune';
\item the instances of `FAW', `FAW Haima' and `FAW Car' with \emph{Haima} models were listed as `FAW Haima';
\item the instances of `FAW', `FAW Hongqi' and `FAW Car' with \emph{Hongqi} models were listed as `FAW Hongqi';
\item `GAC Aion' and `GAC NE' were listed as `GAC Aion';
\item the instances of `GAC' with \emph{Leopaard} models and `GAC Changfeng' were listed as `GAC Changfeng';
\item the instances of `GAC Motor', `Trumpchi' and `GAC' with \emph{Trumpchi} models were listed as `GAC Motor';
\item the instances of `Jiangling Holdings', `JMH' and `Jiangling' with \emph{Landwind} models were listed as `JMH';
\item `Jiangling Motors' and `JMC' were listed as `JMC';
\item the instances of `SAIC Maxus', `SAIC' and `SAIC Motors' with \emph{Maxus} models were listed as `SAIC Maxus';
\item the instances of `SAIC MG', `MG', `MG Motors', `SAIC' and `SAIC Motors' with \emph{MG} models were listed as `SAIC MG';
\item the instances of `SAIC Roewe', `SAIC' and `SAIC Motors' with \emph{Roewe} models were listed as `SAIC Roewe'.
\end{itemize}

Second, we addressed the simultaneous presence of both parent companies and their divisions as manufacturers of the same models (e.g. `Daimler Group' along with `Mercedes-Benz' and `Smart', `GM' along with `Chevrolet' and `Cadillac', etc.); in order to remove such ambiguity, we decided to split the parent companies into their divisions according to the produced models:

\begin{itemize}
\item `Daimler' was split into `Bharat Benz', `Mercedes-Benz', `Mercedes-AMG', `Smart';
\item `Fiat Chrysler' was split into `Abarth', `Alfa Romeo', `Chrysler', `Dodge', `Fiat', `Jeep', `Lancia',  `Ram' and `SRT';
\item `Ford' was split into `Ford' and `Lincoln';
\item `GM' was split into `Buick', `Cadillac', `Chevrolet', `Pontiac', `Opel' and `Opel/Vauxhall';
\item `Jaguar Land Rover' was split into `Jaguar' and `Land Rover'.
\end{itemize}

We would like to stress that the choice of splitting into brands, rather than aggregating into parent companies, was due to the intrinsic arbitrariness of the latter procedure, which requires to set an arbitrary scale to carry out the aggregation itself. As an example, we could have decided to group all the corresponding brands into `Fiat Chrysler' but we could also have decided to move a step further and merge `Fiat Chrysler' with `PSA' (controlling `Citroen', `Peugeot', etc.) into `Stellantis', which is the actual parent company. Moreover, manufacturers that have merged (e.g. `Jaguar' and `Land Rover') have kept assembling different classes of cars, an evidence leading us to suspect that they have kept the relationships with their traditional suppliers, hence can be assumed as representing different entities.\\

Third, we merged pairs of manufacturers in case one of them produced \emph{only} vehicle models also produced by the other one. This led to the following mergers:

\begin{itemize}
\item `Brilliance Jinbei' with `Renault Brilliance Jinbei';
\item `Brilliance Xinyuan' with `Brilliance Shineray';
\item `Chengdu Xindadi' with `Shanghai Maple' and `Geely';
\item `Dongfeng Venucia' with `Dongfeng Nissan';
\item `FAW Xiali' with `Tianjin FAW Xiali';
\item `Guangqi Honda' with `GAC Honda';
\item `Reva' with `Mahindra Reva';
\item `Rongcheng Huatai' with `Hawtai';
\item `Weltmeister' with `WM Motors';
\item `Jiangxi Changhe Suzuki' and `Zhicheng Automobile' with `Changhe Suzuki'.
\end{itemize}

Finally, we merged the following manufacturers as they represent different plants of the same company:

\begin{itemize}
\item `FAW Toyota', `Tianjin FAW Toyota', `SFTM' and `SFTM Changchun Fengyue';
\item `SAIC GM', `SAIC GM Dong Yue' and `SAIC GM Norsom'.
\end{itemize}

\begin{figure*}[h!]
\centering
\includegraphics[width=0.49\linewidth]{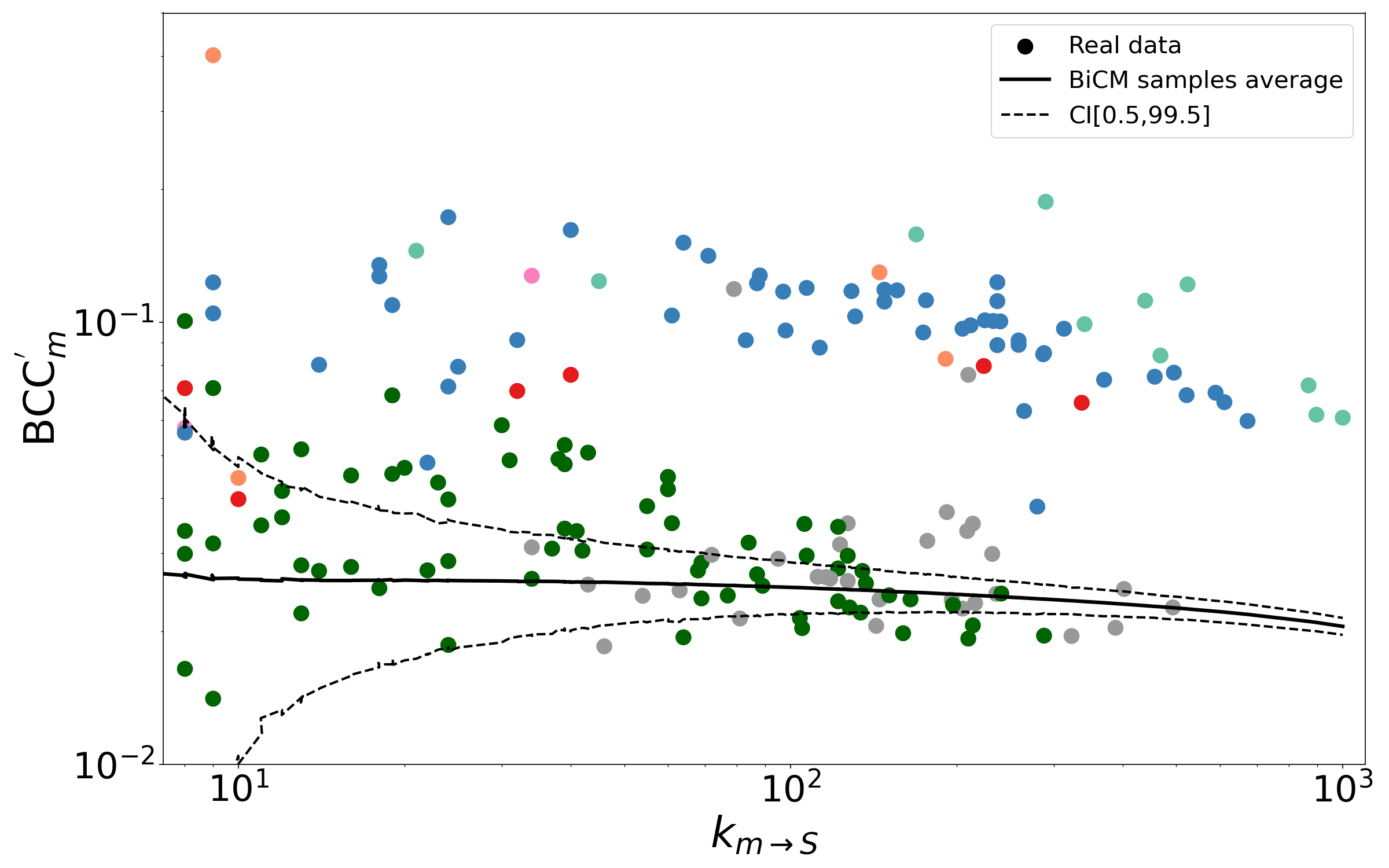}
\includegraphics[width=0.49\linewidth]{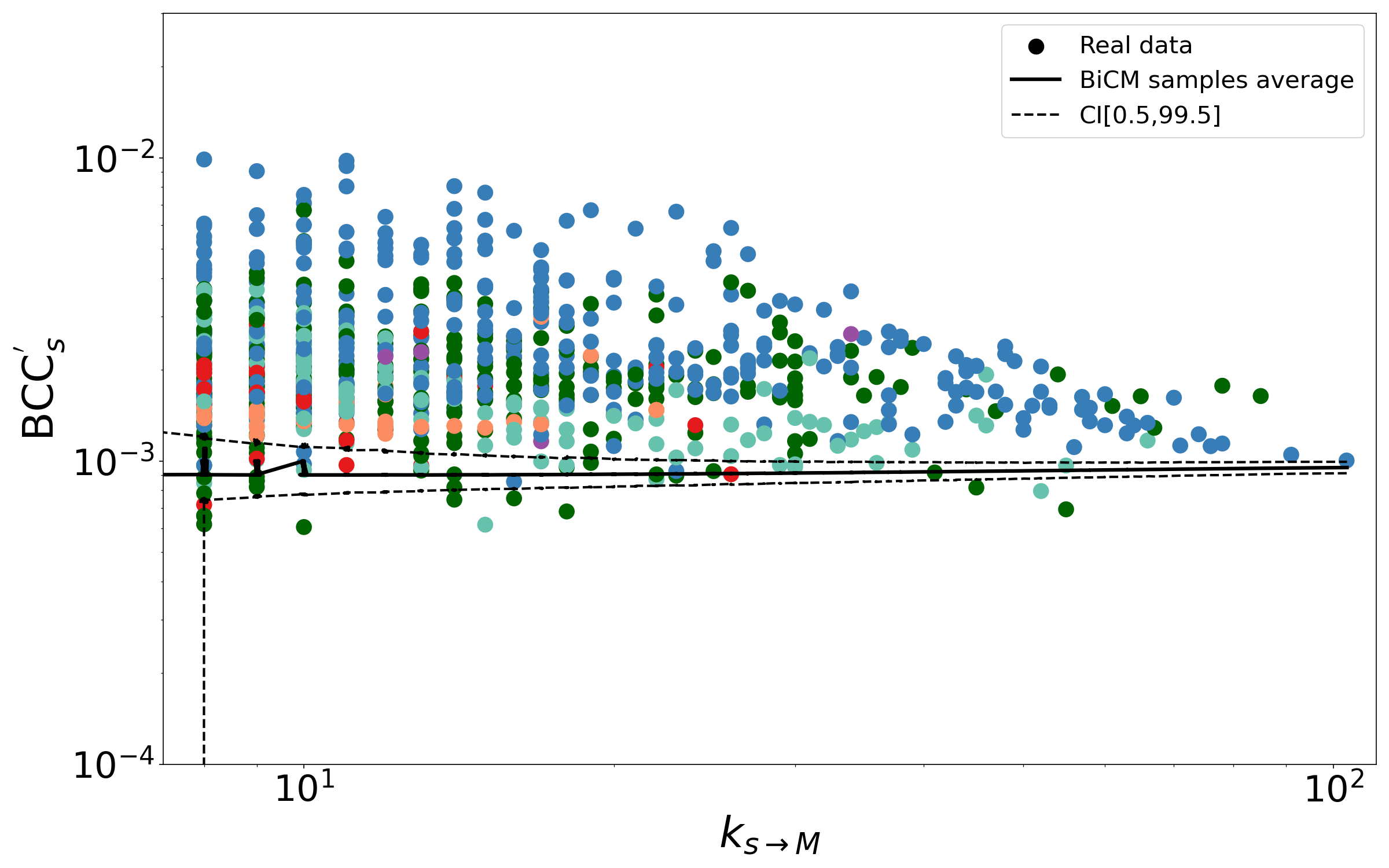}\\
\includegraphics[width=0.49\linewidth]{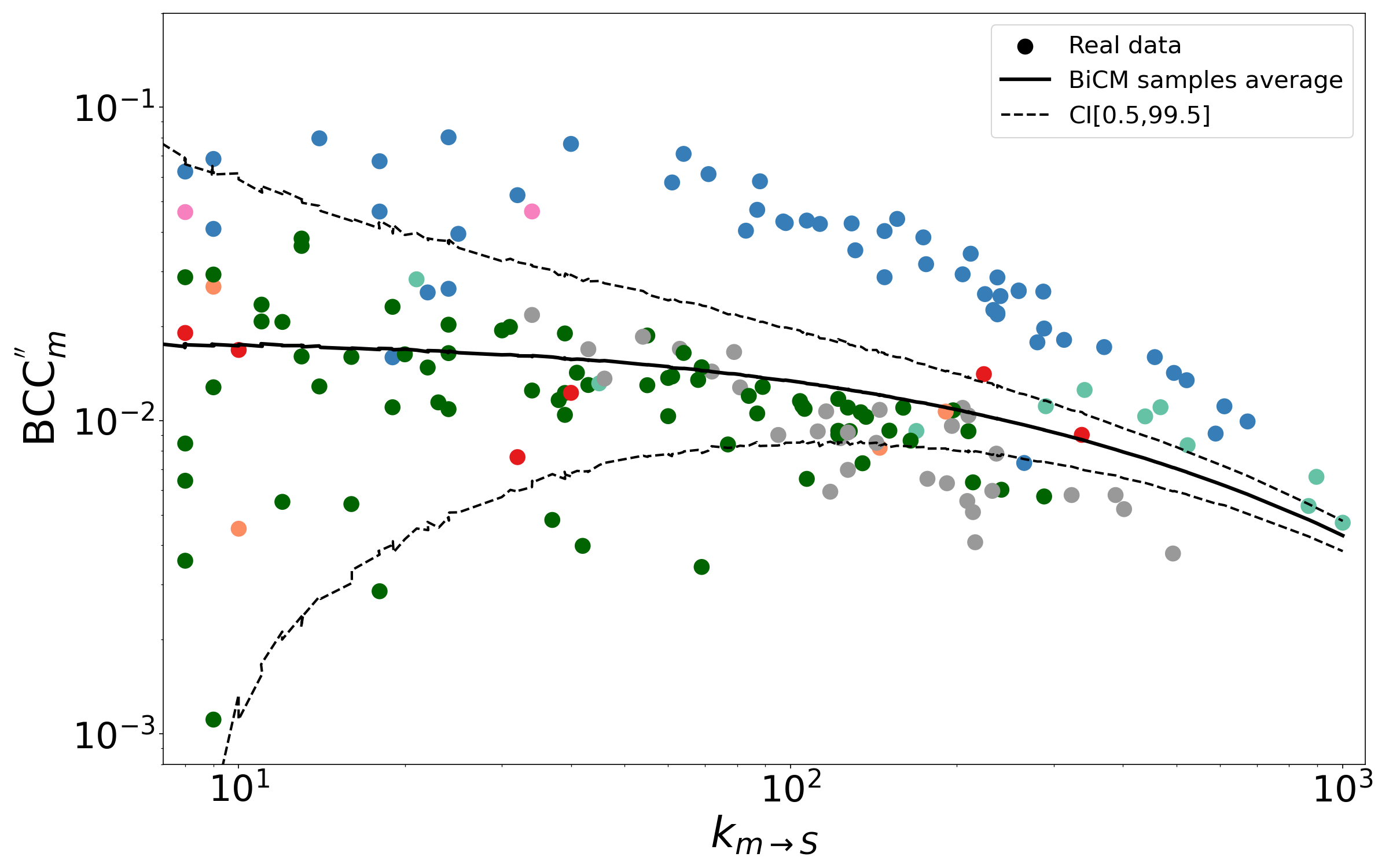}
\includegraphics[width=0.49\linewidth]{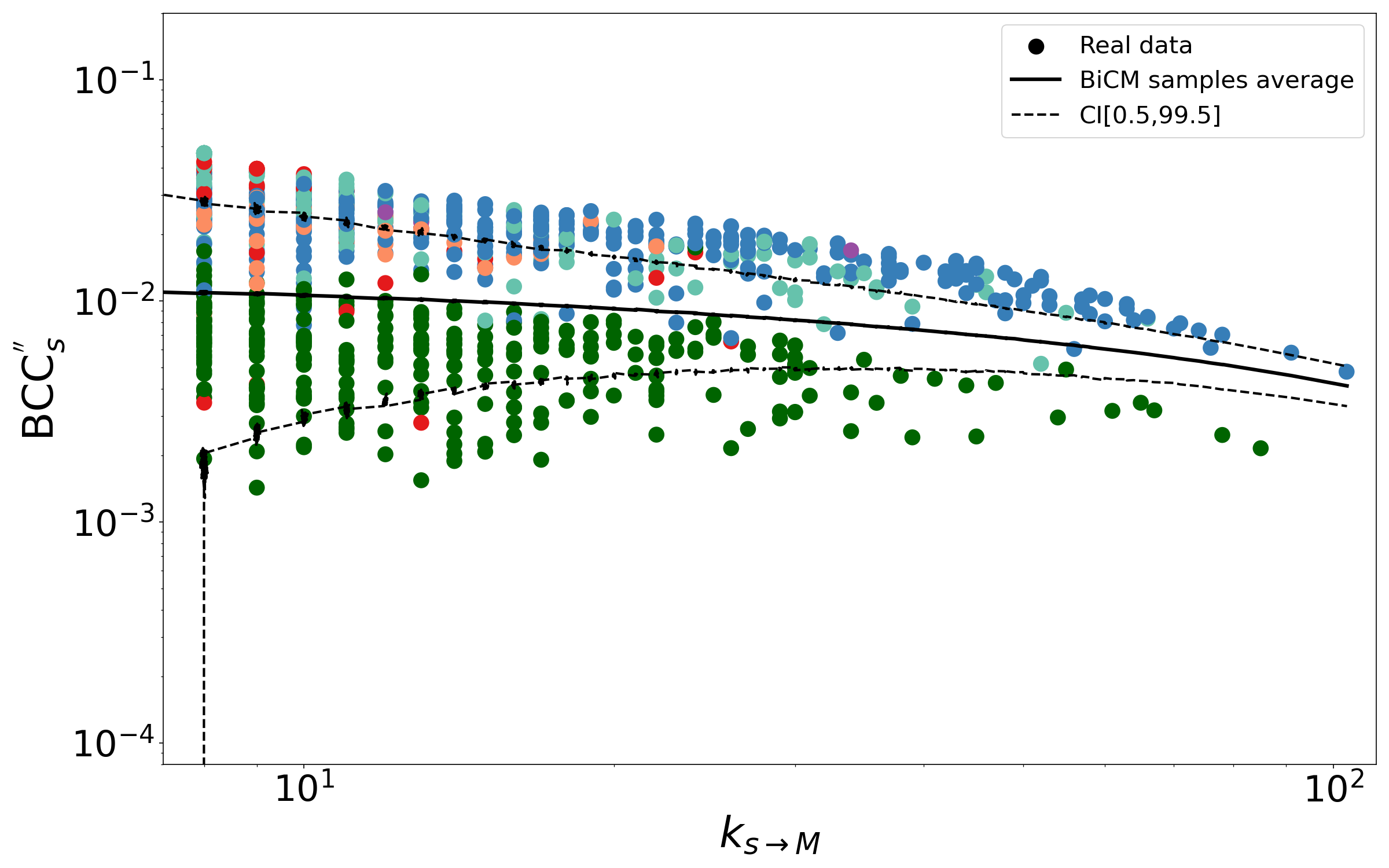}
\caption{Bipartite clustering coefficient (top panels: \eqref{eq:sq_clust_1}; bottom panels: \eqref{eq:sq_clust_3}) scattered versus the degree of manufacturers (left panels) and suppliers (right panels). All trends are overall decreasing, confirming that manufacturers (suppliers) having a large degree are generally closing less squares than manufacturers (suppliers) having a small degree. Again, Western and Japanese firms are clearly distinguished from Chinese firms and JVs on both layers - with definition \eqref{eq:sq_clust_3} providing the neatest separation: while the clustering coefficient of the former is over-represented on both layers, the clustering coefficient of the latter is either in line with the predictions of the BiCM or under-represented. Nodes are colored according to their geographical localization: \textcolor{africa}{$\bullet$} Africa, \textcolor{asean}{$\bullet$} Asean, \textcolor{china}{$\bullet$} Chinese, \textcolor{india}{$\bullet$} Indian, \textcolor{japan}{$\bullet$} Japanese, \textcolor{southam}{$\bullet$} Latin America, \textcolor{middleeast}{$\bullet$} Middle East, \textcolor{russia}{$\bullet$} Russian, \textcolor{west}{$\bullet$} Western and \textcolor{jv}{$\bullet$} Joint Ventures.}
\label{fig12}
\end{figure*}

\section{Appendix B.\\Bipartite clustering coefficient: alternative definitions}

Let us, now, consider two, alternative formulations of the bipartite clustering coefficient. The definition provided in~\cite{lind2005cycles} reads
Let us, now, consider two, alternative formulations of the bipartite clustering coefficient. The definition provided in~\cite{lind2005cycles} reads

\begin{equation}\label{eq:sq_clust_1}
\text{BCC}_m^{'}=\frac{\sum_{s<s'}q_m(s,s')}{\sum_{s<s'}[(u_{s}-\eta_{m}(s,s'))(u_{s'}-\eta_{m}(s,s'))+q_{m}(s,s')]}
\end{equation}
where $\eta_{m}(s,s')=1+q_{m}(s,s')$. According to this definition, the total number of closed squares involving manufacturer $m$ is given by the product of degrees of all pairs of its neighbours, excluding the number of squares that are already closed. Otherwise stated, the total number of closed squares coincides with the number of possible `matchings', achievable by rewiring \emph{existing} links, between the neighbors of $s$ and those of $s'$.

A third, possible definition reads

\begin{equation}\label{eq:sq_clust_3}
\text{BCC}_m^{''}=\frac{\sum_{s<s'}q_m(s,s')}{(M-1)d_m(d_m-1)/2};
\end{equation}
according to it, the total number of closed squares involving manufacturer $m$ coincides with the number of squares that could become closed upon connecting, by adding \emph{new} links, all pairs of its neighbours (e.g. $s$ and $s'$) with the same neighbour.

As shown in figure \ref{fig12}, the obtained trends are very similar to those illustrated in figure \ref{fig5}.

\begin{figure*}[t!]
\centering 
\includegraphics[width=\textwidth]{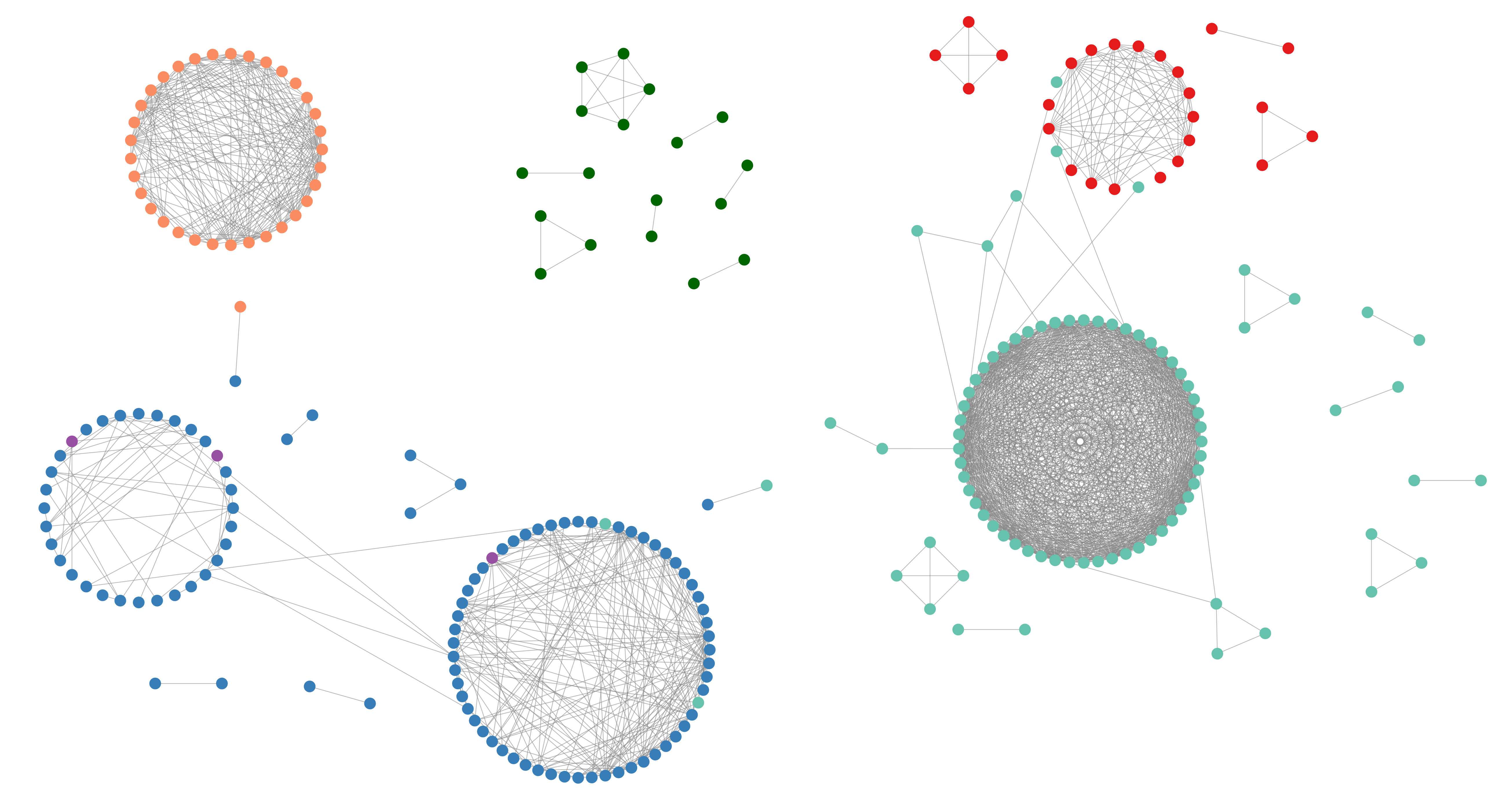}
\hrule
\includegraphics[width=\textwidth]{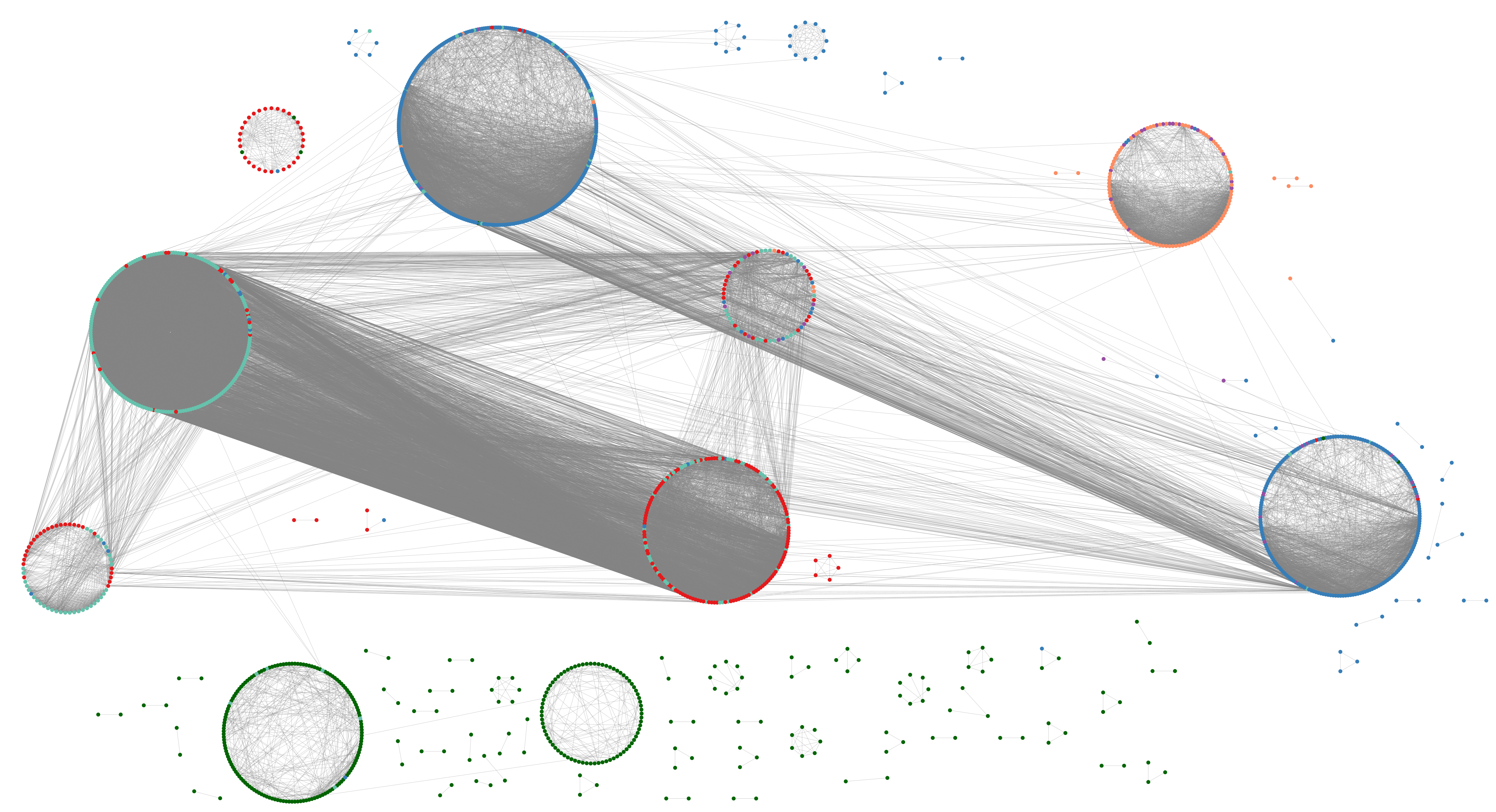}
\caption{Validated projections of suppliers, obtained upon choosing the threshold values $t=10^{-5}$ (top panel) and $t=10^{-3}$ (bottom panel). The two projections both display similar features with respect to the one obtained with a threshold of $t=10^{-4}$ (see fig.\ref{fig9}), such as the fragmentation of Chinese suppliers into several small connected components, the isolation of the Indian cluster, etc. Nodes are colored according to their geographical localization: \textcolor{africa}{$\bullet$} Africa, \textcolor{asean}{$\bullet$} Asean, \textcolor{china}{$\bullet$} Chinese, \textcolor{india}{$\bullet$} Indian, \textcolor{japan}{$\bullet$} Japanese, \textcolor{southam}{$\bullet$} Latin America, \textcolor{middleeast}{$\bullet$} Middle East, \textcolor{russia}{$\bullet$} Russian, \textcolor{west}{$\bullet$} Western.
}
\label{fig13}
\end{figure*}

\section{Appendix C.\\Choosing the threshold for the validated projection of suppliers}

When building the validated network of suppliers, the application of the FDR criterion as described in the main text yields an empty graph. In order to retain some information, we softened the validation procedure by neglecting the correction for testing multiple hypotheses while lowering the threshold $t$ - for the sake of robustness, we employed three, different values, i.e. $t=10^{-3}$, $t=10^{-4}$ and $t=10^{-5}$. The corresponding projections (i.e. figure \ref{fig9}, obtained upon choosing $t=10^{-4}$, and figure \ref{fig13}, obtained upon choosing $t=10^{-3}$ and $t=10^{-5}$) show comparable structural features - namely the fragmentation of Chinese suppliers into several small connected components, the isolation of the Indian cluster, the division of Western firms into one American and one European cluster, etc. - thus corroborating the validity of these findings.

\clearpage

\section{Appendix D.\\Composition of communities of products in the monopartite projection}

\begin{table*}[h!]
\caption{\label{tab3}List of products populating the corresponding projection, together with their technological classification and the community they belong to. As stressed in the main text, the community structure we identify does not match with the (technological) taxonomy provided by the MarkLines platform; rather, it identifies the different car systems.}
\begin{ruledtabular}
\begin{tabular}{lll}
Product & Layer & Community \\
\hline
Boot & Interior/Exterior & Cooling System \\
Bush & Interior/Exterior & Cooling System \\
Bearing & Interior/Exterior & Cooling System \\
Fuel hose & Powertrain & Cooling System \\
Engine mount & Powertrain & Cooling System \\
Radiator & Powertrain & Cooling System \\
Radiator hose & Powertrain & Cooling System \\
Fuel line & Powertrain & Cooling System \\
Oil cooler & Powertrain & Cooling System \\
Cooling fan module & Powertrain & Cooling System \\
Engine bearing & Powertrain & Cooling System \\
Engine cooling module & Powertrain & Cooling System \\
Inter cooler & Powertrain & Cooling System \\
Power steering hose & Chassis/Body & Cooling System \\
Wheel bearing & Chassis/Body & Cooling System \\
Brake hose & Chassis/Body & Cooling System \\
Brake line & Chassis/Body & Cooling System \\
Heater hose & Electrical & Cooling System \\
Radiator fan controller & Electrical & Cooling System \\
AC HVAC & Electrical & Cooling System \\
Condenser & Electrical & Cooling System \\
AC hose & Electrical & Cooling System \\
Heater & Electrical & Cooling System \\
Air conditioner ECU & Electrical & Cooling System \\
AC compressor & Electrical & Cooling System \\
\hline
Inside door handle & General Parts & Doors \\ Outside door handle & General Parts & Doors \\
Shift lever & Powertrain & Doors \\
Side door closure & Chassis/Body & Doors \\
Convertible roof & Chassis/Body & Doors \\
Key cylinder steering lock & Chassis/Body & Doors \\
Back door/trunk lock & Chassis/Body & Doors \\
Door module & Chassis/Body & Doors \\
Window regulator & Chassis/Body & Doors \\
Side door lock & Chassis/Body & Doors \\
Wiper system & Chassis/Body & Doors \\
Wiper arm/blade & Chassis/Body & Doors \\
Hinge & Chassis/Body & Doors \\
Tailgate trunk closure & Chassis/Body & Doors \\
Lever combination switch & Electrical & Doors \\
Power tailgate trunk ECU & Electrical & Doors \\
Junction box & Electrical & Doors \\
Power sliding door ECU & Electrical & Doors \\
Horn & Electrical & Doors \\
Relay fuse & Electrical & Doors \\
Wiring harness & Electrical & Doors \\
Electrical connector & Electrical & Doors \\
\end{tabular}
\end{ruledtabular}
\end{table*}

\begin{table*}[t!]
\begin{ruledtabular}
\begin{tabular}{lll}
Product & Layer & Community \\
\hline
Motor actuator & Electrical & Doors \\
Switch & Electrical & Doors \\
Keyless entry start system & Electrical & Doors \\
Pedestrian protection airbag & General Parts & Electronic Control Units \\
HVEV ECU & Powertrain & Electronic Control Units \\
Ignition coil & Powertrain & Electronic Control Units \\
Diesel injector & Powertrain & Electronic Control Units \\
Engine management system & Powertrain & Electronic Control Units \\
Battery control ECU & Powertrain & Electronic Control Units \\
On board charger & Powertrain & Electronic Control Units \\
Inverter & Powertrain & Electronic Control Units \\
DC DC converter & Powertrain & Electronic Control Units \\
Battery & Powertrain & Electronic Control Units \\
Fuel pump &  Powertrain & Electronic Control Units \\
Glow plug & Powertrain & Electronic Control Units \\
Spark plug & Powertrain & Electronic Control Units \\
Starter motor &  Powertrain & Electronic Control Units \\
Alternator generator & Powertrain & Electronic Control Units \\
Throttle body & Powertrain & Electronic Control Units \\
Fuel injector & Powertrain & Electronic Control Units \\
HVPHVEV battery & Powertrain & Electronic Control Units \\
Power steering motor & Chassis/Body & Electronic Control Units \\
Active engine mount ECU & Electrical & Electronic Control Units \\
 Seatbelt ECU & Electrical & Electronic Control Units \\
Stop start system ECU & Electrical & Electronic Control Units \\
Air flow sensor & Electrical & Electronic Control Units \\
Lane keeping assist system ECU & Electrical & Electronic Control Units \\
Electronically controlled all wheel drive ECU & Electrical & Electronic Control Units \\
Cruise control & Electrical & Electronic Control Units \\
Camera ECU & Electrical & Electronic Control Units \\
Door ECU & Electrical & Electronic Control Units \\
Clearance sonar ECU & Electrical & Electronic Control Units \\
Head lamp ECU & Electrical & Electronic Control Units \\
Head lamp leveling ECU & Electrical & Electronic Control Units \\
Power seat ECU & Electrical & Electronic Control Units \\
Occupant detection system & Electrical & Electronic Control Units \\
Object detection ECU & Electrical & Electronic Control Units \\
Antitheft immobilizer & Electrical & Electronic Control Units \\
Power steering ECU & Electrical & Electronic Control Units \\
Milliwave and laser radar & Electrical & Electronic Control Units \\
Steering sensor & Electrical & Electronic Control Units \\
Onboard camera & Electrical & Electronic Control Units \\
Pedal sensor & Electrical & Electronic Control Units \\
Head up display & Electrical & Electronic Control Units \\
Pressure sensor & Electrical & Electronic Control Units \\
Temperature sensor & Electrical & Electronic Control Units \\
Knock sensor & Electrical & Electronic Control Units \\
Glowplug controller & Electrical & Electronic Control Units \\
Airbag sensor & Electrical & Electronic Control Units \\
ADAS ECU & Electrical & Electronic Control Units \\
Electronic control unit & Electrical & Electronic Control Units \\
Rain light sensor & Electrical & Electronic Control Units \\
Car navigation system & Electrical & Electronic Control Units \\
Body control ECU & Electrical & Electronic Control Units \\
Display & Electrical & Electronic Control Units \\
ABS ECU & Electrical & Electronic Control Units \\
Tire pressure monitoring system ECU & Electrical & Electronic Control Units \\
OBD interface & Electrical & Electronic Control Units \\
Park assist system & Electrical & Electronic Control Units \\
Airbag ECU & Electrical & Electronic Control Units \\
Oxygen sensor & Electrical & Electronic Control Units \\
\end{tabular}
\end{ruledtabular}
\end{table*}

\begin{table*}[t!]
\begin{ruledtabular}
\begin{tabular}{lll}
Product & Layer & Community \\
\hline
Crank cam sensor & Electrical & Electronic Control Units \\
Transmission ECU & Electrical & Electronic Control Units \\
Electronically controlled suspension ECU & Electrical & Electronic Control Units \\
Antenna & Electrical & Electronic Control Units \\
In-vehicle infotainment & Electrical & Electronic Control Units \\
Speed sensor & Electrical & Electronic Control Units \\
Telematics & Electrical & Electronic Control Units \\
Meter & Electrical & Electronic Control Units \\
Car audio & Electrical & Electronic Control Units \\
Engine control unit & Electrical & Electronic Control Units \\
Seatbelt pretensioner & General Parts & Airbags, Wheel \\
Side airbag & General Parts & Airbags, Wheel \\
Driver airbag & General Parts & Airbags, Wheel \\
Knee airbag & General Parts & Airbags, Wheel \\
Passenger airbag & General Parts & Airbags, Wheel \\
Curtain airbag & General Parts & Airbags, Wheel \\
Seatbelt & General Parts & Airbags, Wheel \\
Steering wheel & Chassis/Body & Airbags, Wheel \\
\hline
Valve spring & Powertrain & Steering System \\
Rack end & Chassis/Body & Steering System \\
Power steering assist unit & Chassis/Body & Steering System \\
Tie rod end & Chassis/Body & Steering System \\
Suspension ball joint & Chassis/Body & Steering System \\
Stabilizer & Chassis/Body & Steering System \\
Power steering pump & Chassis/Body & Steering System \\
Steering gear & Chassis/Body & Steering System \\
Steering column & Chassis/Body & Steering System \\
Suspension spring & Chassis/Body & Steering System \\
Shock absorber & Chassis/Body & Steering System \\
Steering system & Chassis/Body & Steering System \\
\hline
Clutch master cylinder & Powertrain & Breaking System \\
Steering knuckle & Chassis/Body & Breaking System \\
Drum brake shoe & Chassis/Body & Breaking System \\
Brake wheel cylinder & Chassis/Body & Breaking System \\
Drum brake lining & Chassis/Body & Breaking System \\
Brake master cylinder & Chassis/Body & Breaking System \\
Drum brake & Chassis/Body & Breaking System \\
Brake booster & Chassis/Body & Breaking System \\
Parking brake & Chassis/Body & Breaking System \\
Corner module & Chassis/Body & Breaking System \\
Disc brake pad & Chassis/Body & Breaking System \\
ABS ESC & Chassis/Body & Breaking System \\
Disc brake caliper & Chassis/Body & Breaking System \\
Brake disc rotor & Chassis/Body & Breaking System \\
Electric park brake ECU & Electrical & Breaking System \\
Vehicle dynamics control & Electrical & Breaking System \\
\hline
Fuel filter & Powertrain & Engine \\
Transmission seal & Powertrain & Engine \\
V belt & Powertrain & Engine \\
Exhaust manifold gasket & Powertrain & Engine \\
Engine ass Y & Powertrain & Engine \\
EGR system & Powertrain & Engine \\
Piston pin & Powertrain & Engine \\
Timing belt/chain & Powertrain & Engine \\
Cylinder head & Powertrain & Engine \\
Cylinder head gasket & Powertrain & Engine \\
Air intake module & Powertrain & Engine \\
Connecting rod & Powertrain & Engine \\
Engine valve & Powertrain & Engine \\
Camshaft & Powertrain & Engine \\
Intake manifold & Powertrain & Engine \\
\end{tabular}
\end{ruledtabular}
\end{table*}

\begin{table*}[t!]
\begin{ruledtabular}
\begin{tabular}{lll}
Product & Layer & Community \\
\hline
Cylinder head cover & Powertrain & Engine \\
Crankshaft & Powertrain & Engine \\
Piston ring & Powertrain & Engine \\
Timing system & Powertrain & Engine \\
Cylinder liner & Powertrain & Engine \\
Oil filter & Powertrain & Engine \\
Piston & Powertrain & Engine \\
Valve guide/seat & Powertrain & Engine \\
Carbon canister & Powertrain & Engine \\
Cylinder block & Powertrain & Engine \\
Oil pump & Powertrain & Engine \\
Water pump & Powertrain & Engine \\
Air cleaner/filter & Powertrain & Engine \\
Heat shield & Chassis/Body & Engine \\
Cabin air filter & Electrical & Engine \\
Window glass & General Parts & Windows \\
Sunroof & Chassis/Body & Windows \\
\hline
Exhaust manifold & Powertrain & Exhaust System \\
Fuel supply system module & Powertrain & Exhaust System \\
Diesel particulate filter & Powertrain & Exhaust System \\
Catalytic converter & Powertrain & Exhaust System \\
Muffler & Powertrain & Exhaust System \\
Exhaust system & Powertrain & Exhaust System \\
Fuel tank & Chassis/Body & Exhaust System \\
Fuel filler & Chassis/Body & Exhaust System \\
\hline
Head lamp cleaner & General Parts & Lighting System \\
High mounted stop lamp & General Parts & Lighting System \\
Interior mirror & General Parts & Lighting System \\
Fog lamp & General Parts & Lighting System \\
Head lamp (AFS) & General Parts & Lighting System \\
License plate lamp & General Parts & Lighting System \\
Interior lighting & General Parts & Lighting System \\
Exterior mirror & General Parts & Lighting System \\
Head lamp & General Parts & Lighting System \\
Rear lamp & General Parts & Lighting System \\
Window washer & Chassis/Body & Lighting System \\
Adaptive front-lighting system ECU & Electrical & Lighting System \\
\hline
Glass run channel & General Parts & Cabin \\
Sun visor & General Parts & Cabin \\
Roof rail & General Parts & Cabin \\
Trunk tailgate trim & General Parts & Cabin \\
Molding & General Parts & Cabin \\
Floor mat & General Parts & Cabin \\
Seat frame & General Parts & Cabin \\
Cockpit module & General Parts & Cabin \\
Floor carpet & General Parts & Cabin \\
Cup holder & General Parts & Cabin \\
Spoiler & General Parts & Cabin \\
Headrest & General Parts & Cabin \\
Wheel cover cap & General Parts & Cabin \\
Pedal & General Parts & Cabin \\
Pedal box & General Parts & Cabin \\
Emblem & General Parts & Cabin \\
Glove box & General Parts & Cabin \\
Body side molding & General Parts & Cabin \\
Door panel & General Parts & Cabin \\
Weatherstrip & General Parts & Cabin \\
Bumper & General Parts & Cabin \\
Seat trim & General Parts & Cabin \\
Seat lumbar support & General Parts & Cabin \\
Seat adjuster recliner & General Parts & Cabin \\
\end{tabular}
\end{ruledtabular}
\end{table*}

\begin{table*}[t!]
\begin{ruledtabular}
\begin{tabular}{lll}
Product & Layer & Community \\
\hline
Radiator Grille & General Parts & Cabin \\
Headliner & General Parts & Cabin \\
Seat & General Parts & Cabin \\
Door trim & General Parts & Cabin \\
Instrument panel & General Parts & Cabin \\
Console & General Parts & Cabin \\
Duct & Interior/Exterior & Cabin \\
Acoustic insulator & Interior/Exterior & Cabin \\
Axle & Powertrain & Cabin \\
Subframe suspension member & Chassis/Body & Cabin \\
Door frame & Chassis/Body & Cabin \\
Bumper beam & Chassis/Body & Cabin \\
Chassis frame & Chassis/Body & Cabin \\
Cross car beam & Chassis/Body & Cabin \\
Suspension control arm & Chassis/Body & Cabin \\
Side impact beam & Chassis/Body & Cabin \\
Suspension module & Chassis/Body & Cabin \\
Crossmember & Chassis/Body & Cabin \\
Front end module & Chassis/Body & Cabin \\
Transfer & Powertrain & Clutch \\
Electric all wheel drive motor & Powertrain & Clutch \\
CVT & Powertrain & Clutch \\
All wheel drive & Powertrain & Clutch \\
Automated manual transmission & Powertrain & Clutch \\
Reduction gear for EV & Powertrain & Clutch \\
Clutch slave cylinder & Powertrain & Clutch \\
Propshaft & Powertrain & Clutch \\
Clutch & Powertrain & Clutch \\
Flywheel & Powertrain & Clutch \\
Torque converter & Powertrain & Clutch \\
Differential & Powertrain & Clutch \\
Manual transmission & Powertrain & Clutch \\
Dual clutch transmission & Powertrain & Clutch \\
Clutch disc & Powertrain & Clutch \\
Drive shaft & Powertrain & Clutch \\
Vehicle control unit & Powertrain & Clutch \\
Traction motor & Powertrain & Clutch \\
Automatic transmission & Powertrain & Clutch \\
e4WD ECU & Electrical & Clutch \\
\end{tabular}
\end{ruledtabular}
\end{table*}

\clearpage

\begin{figure}[t!]
\centering
\includegraphics[width=\textwidth]{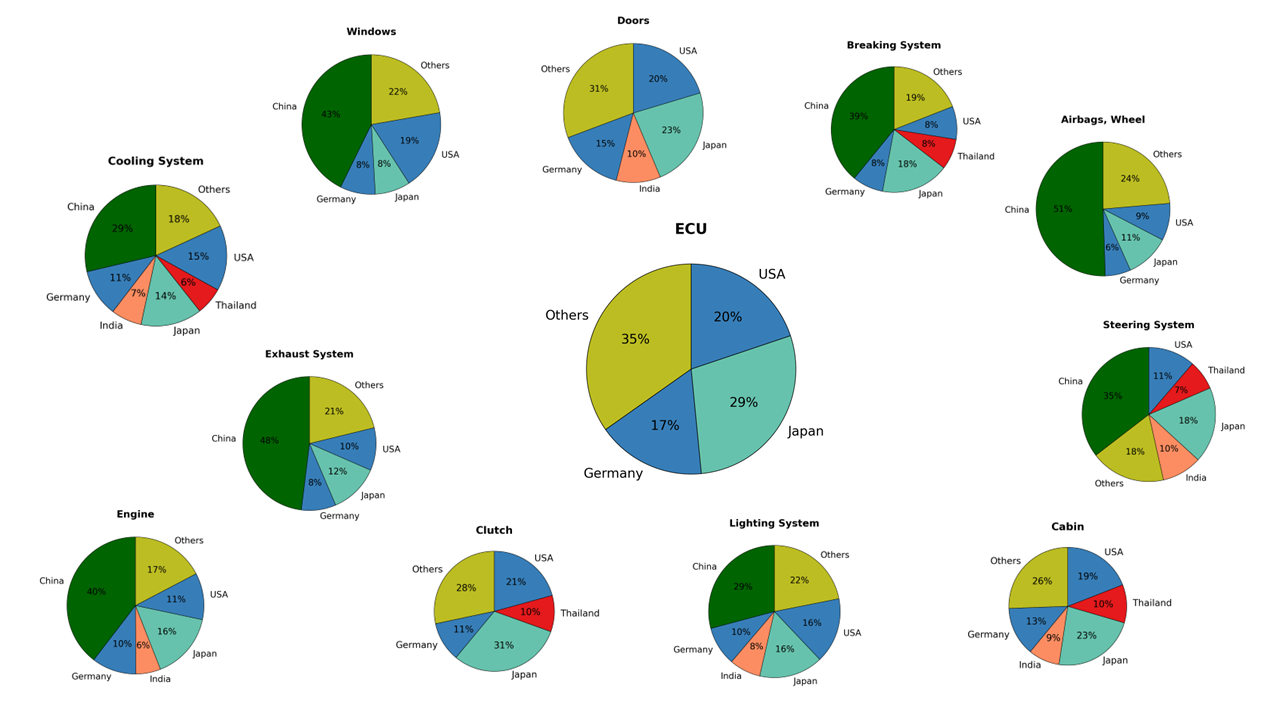}
\caption{Pie charts illustrating the geographical origin of the suppliers selling the products populating the communities (of the monopartite projection) of Figure \ref{fig10}: only countries with a share bigger than $5\%$ are explicitly shown, the others having being grouped into \emph{Others}. China, Germany, Japan and USA are the most represented countries. China is extremely active in some sectors and completely absent in others while the other, three ones have a similar share in all communities.}
\label{fig14}
\end{figure}

\section{Appendix E.\\Geographical characterization of communities of products}

{As we have already observed, suppliers tend to focus their production on `coherent' sets of products. Inspecting the nationality of the suppliers of these products leads to the pie charts depicted in Figure \ref{fig14}.}

\end{document}